\documentclass[fleqn,usenatbib]{mnras}
\pdfoutput=1
\usepackage{newtxtext,newtxmath}

\usepackage[T1]{fontenc}
\usepackage{ae,aecompl}

\usepackage{comment}

\usepackage{graphicx}	
\usepackage{amsmath}	
\let\oldAA\AA 
\renewcommand{\AA}{\text{\normalfont\oldAA}} 
\usepackage{float} 

\newcommand{\imw}{$i$--$W3$}

\newcommand{\rmwo}{$r$--$W1$}

\newcommand{\rmz}{$r$--$z$}

\newcommand{\kms}{km~s$^{-1}$}

\newcommand{\ledd}{$L_{\text{Edd}}$}
\newcommand{\ergs}{erg s$^{-1}$}
\newcommand{\mum}{$\mu$m}
\newcommand{\lam}{$\lambda$}
\newcommand{\Lam}{$\Lambda$}

\newcommand{\hb}{H$\beta$}

\newcommand{\heii}{\mbox{He\,{\sc ii}}}

\newcommand{\civ}{\mbox{C\,{\sc iv}}}
\newcommand{\ciii}{\mbox{C\,{\sc iii}}}
\newcommand{\cii}{\mbox{C\,{\sc ii}}}

\newcommand{\siiv}{\mbox{Si\,{\sc iv}}}
\newcommand{\siiii}{\mbox{Si\,{\sc iii}}}
\newcommand{\siii}{\mbox{Si\,{\sc ii}}}
\newcommand{\siv}{\mbox{S\,{\sc iv}}}

\newcommand{\nv}{\mbox{N\,{\sc v}}}

\newcommand{\oiii}{\mbox{O\,{\sc iii}}}
\newcommand{\oii}{\mbox{O\,{\sc ii}}}
\newcommand{\oi}{\mbox{O\,{\sc i}}}

\newcommand{\feii}{\mbox{Fe\,{\sc ii}}}

\newcommand{\mgii}{\mbox{Mg\,{\sc ii}}}

\newcommand{\aliii}{\mbox{Al\,{\sc iii}}}

\title[\civ\ Outflows]{BOSS Quasar Outflows Traced by \civ}
\author[Gillette et al.]{
Jarred Gillette,$^{1}$\thanks{E-mail: jgill016@ucr.edu (JOG)}
Fred Hamann$^{1}$
\\
$^{1}$Department of Physics \& Astronomy, University of California Riverside, 900 University Avenue, Riverside CA 92521, USA
}

\date{Accepted XXX. Received YYY; in original form ZZZ}

\pubyear{2023}

\begin{document}
\label{firstpage}
\pagerange{\pageref{firstpage}--\pageref{lastpage}}
\maketitle

\begin{abstract}
We investigate possible factors that drive fast quasar outflows using a sample of 39,249 quasars at median redshift $\langle z \rangle \approx$ 2.17. Unique to this study, the quasar redshifts are re-measured based on the \mgii\ emission line, allowing for exploration of unprecedented outflow velocities (>6000 \kms) while maintaining statistical significance and uniformity. We measure reliable \civ\ blueshifts for 1178 quasars with velocities >2500 \kms. From those, 255(13) quasars have blueshifts above 4000(6000) \kms, with the highest \civ\ velocity $\approx$ 7000 \kms. Several correlations are observed, where higher \civ\ blueshifts in general are in quasars with broader, weaker \civ\ emission profiles, weak \heii\ emission, larger Eddington ratios, and bluer UV continuum slope across the rest-frame UV to Near-IR. Analysis reveals two primary factors contributing to faster outflows: higher Eddington ratios, and softer far-UV continuum (h$\nu$ >24.6 eV). We find supporting evidence that radiative line-driving may generate extreme outflow velocities, influenced by multiple factors as suggested by the aforementioned correlations. This evidence highlights the importance of considering a multi-dimensional parameter space in future studies when analyzing large \civ\ blueshifts to determine the fundamental causes of outflows.
\end{abstract}

\begin{keywords}
galaxies: active -- quasars: emission lines -- quasars: general -- galaxies: high-redshift
\end{keywords}



\section{Introduction}
\label{sec:sec_intro}

High-z quasars are associated with formation and growth of massive galaxies, and it is believed that the feedback from these quasars plays a crucial role in regulating both galaxy growth and star formation processes \citep{Hopkins+08,Hopkins+16,WylezalekZakamska2016,Veilleux+16,RupkeGultekinVeilleux17,Baron+17,Vayner+21}. However, the physical nature and driving mechanism(s) of quasar outflows remain poorly understood. We need, in particular, better empirical constraints on the quasar properties and physical conditions that correlate with a higher incidence, greater speeds, or more powerful outflows. 

Quasar outflows are often studied via blueshifted broad absorption lines (BALs) and their narrower cousins, mini-BALs, in rest-frame UV spectra \citep{Weymann+91,Hamann+04,Baskin+13,Muzahid+13,Hamann+19,Chen+19}. Another approach is to use the blueshifts in their broad emission-lines (BELs). These strategies are complementary, but the BELs offer unique advantages. One is that the location of BEL regions is known from reverb studies to be unambiguously near the central engines, at radial distances of roughly 0.01 to 0.1 pc in luminous quasars \citep[][and references therein]{Netzer_20}. Thus, emission lines provide specific constraints on quasar outflow models. Another advantage is that emission-line studies are less sensitive to orientation and time-dependent effects than absorption lines, which can only measure the gas/outflow structures that happen to lie along our line of sight to the continuum source during the observing epoch. Specific values of emission blueshift can vary depending on the sample and selection criteria, previous studies have found typical blueshifts ranging from a few thousand \kms\ in BELs, to tens of thousands of \kms\ for BALs \citep[e.g.,][]{Richards+11,Coatman+19, Temple+20, Rankine+20}.

We do not know the geometry with certainty, but strongly blueshifted profiles from high ionization lines suggest the emission comes from a wind that is moving toward us, with the receding component blocked by the optically thick accretion disk. Intermediate and lower ionization lines may be emitted by gas in the accretion disk atmosphere, or a low-velocity base of the wind \citep{Leighly+04a,Leighly04b,Casebeer+06,Richards+11,Hamann+17,Rankine+20,Temple+23}.

Investigating these line shifts require accurate systemic redshift of the quasar. In many quasar blueshift studies the systemic redshift is estimated by fitting of multiple emission lines simultaneously, which are assumed to be at or near their rest wavelength \citep{Richards+11,Coatman+16,Coatman+19,Rankine+20,Temple+20,Temple+23}. Many studies estimate redshift using emission-lines or emission-templates that may include higher ion lines. High ion lines like \civ\ \lam1549 can be blueshifted from systemic by several hundreds of \kms\ \citep{Gaskell82,Richards+02}. The best indicators are narrow emission lines arising from the extended host galaxies and halo environments. Efforts have been undertaken to establish accurate redshifts for select quasar populations in order to better constrain extreme outflow velocities \citep[e.g.,][]{Gillette+23b}.

Low-ion BELs such as \mgii, \oi\ \lam1304, and \cii\ \lam1335 are also generally good indicators of quasar redshifts, e.g., with small or negligible offsets relative to narrow [\oii] and [\oiii] emission lines. \mgii\ is estimated to be near the systemic redshift when compared to these other lines in high redshift quasars \citep{Richards+02}, and is much more reliable than high-ion lines like \civ. A mean shift between \mgii\ - \civ\ has been shown of $\sim$920 $\pm$ 750 \kms\ \citep[Ref. appendix in,][]{Shen+07}.

For this work, we present a sample of \mgii\ emission-line measurements for 39,249 quasars from BOSS DR12. We then use these data to study the blueshifts in the \civ\ BEL compared to a wide variety of other quasar properties that can be measured or derived from this dataset. This study differs from previous work by determining redshifts only from careful fitting of the \mgii\ line, allowing us to explore an unprecedented range of outflow velocities uniformly, and maintain a statistically significant sample size. We can then focus more sharply on trends in other lines and quasar properties that correlate with blueshifts.

In this paper we analyze blueshifted emissions of \civ, and compare their emission profiles and UV features to understand what correlates with strong outflows. It is organized as follows. Section 2 describes criteria for selecting quasars from catalogue spectra. Section 3 describes our methods and measurements of blueshift, \mgii\ line features, and black hole mass. Section 4 describes correlations in measured parameters of UV emission features, such as blueshift, emission line widths, Eddington ratio, and spectral shape. Section 5 we discuss implications for quasar outflows. Section 6 concludes the paper. Throughout this paper we adopt a \Lam-CDM cosmology with $H_0$~=~69.6~\kms~Mpc$^{-1}$, $\Omega_\text{M}$~=~0.286 and $\Omega_\Lambda$~=~0.714, as adopted by the online cosmology calculator developed by \citet{Wright06}. All magnitudes are on the AB system. Reported wavelengths are in vacuum and in the heliocentric frame.

\section{Quasar Samples and Data Sets}\label{sec_data_set}

\begin{figure}
    \includegraphics[width=\columnwidth]{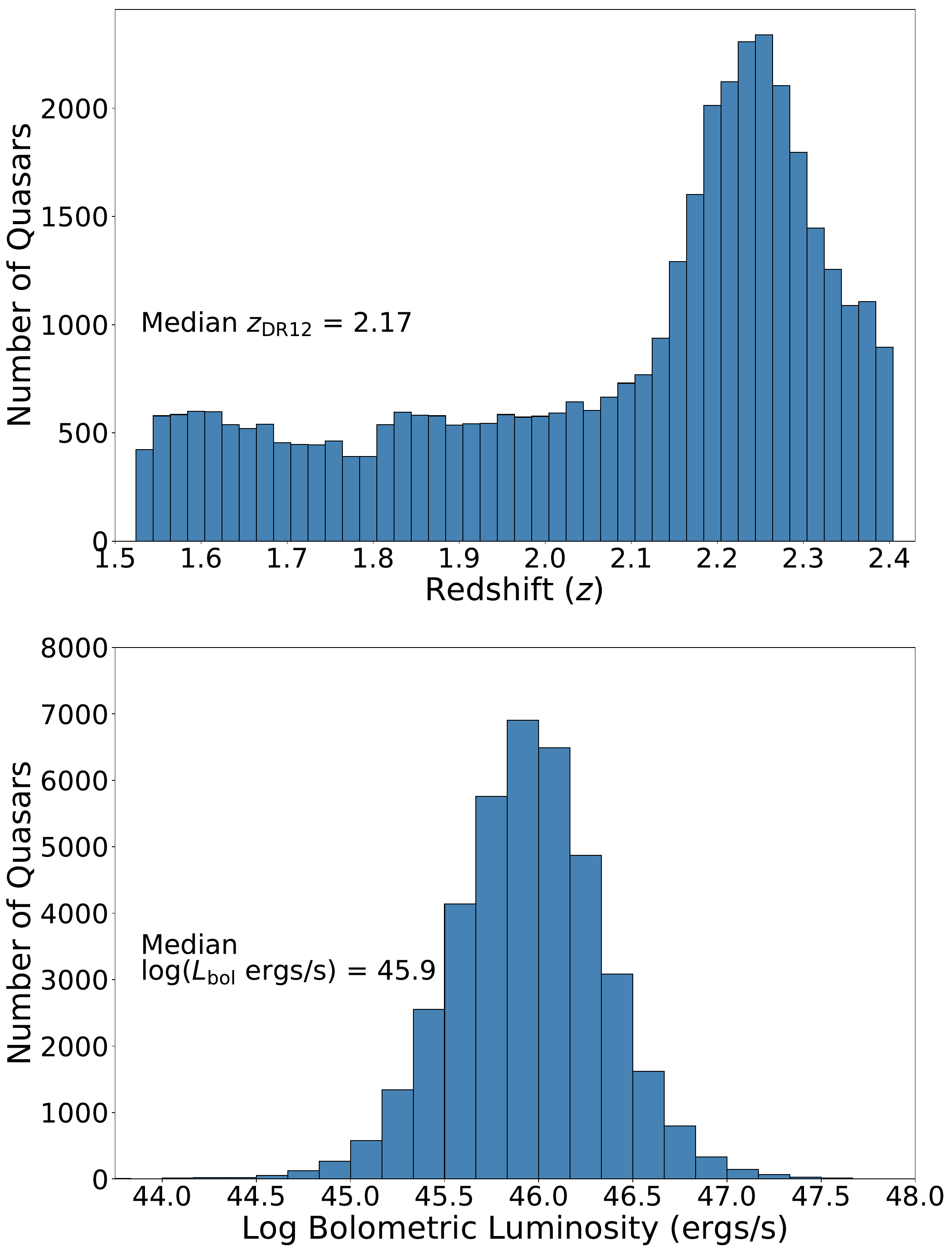}
    \caption{Redshifts and computed Bolometric luminosities for our full sample. \textit{Top:} Redshift distribution as provided from DR12Q. \textit{Bottom:} Bolometric luminosities computed from monochromatic luminosities as described in Section \ref{sec:sec_VBHM}. }
    \label{fig:fig_figure1}
\end{figure}

We started from a \civ\ emission-line sample of 173,636 quasars, measured by \citep{Hamann+17}, from the Baryon Oscillations Survey (BOSS) from the Sloan Digital Sky Survey Data Release 12 \citep[hereafter DR12Q;][]{Paris2014,Paris2017}. We limit our study to Type 1 sources based on the crude distinction FWHM(\civ) $\geq$ 2000 km s$^{-1}$ following previous studies by \citet{Alexandroff13} and \citet{Ross15}. This was done to avoid contamination by Type 2 quasars unless \civ\ FWHM is explicitly used for comparison to other parameters. Sample selection criteria and quasar totals are summarized in Table \ref{tab:tab_table1}. Other subsamples after the ``full'' sample use additional constraints as described in the sections noted. Figure \ref{fig:fig_figure1} shows redshift and luminosity distributions. To ensure \civ\ and \mgii\ spectral coverage we are limited to the redshift range 1.50 $\leq z \leq$ 2.46. By our quality standards, outlines in Section \ref{sec:sec_cont_line_fit}, we have 39,249 quasars with a median redshift of $\langle z \rangle \approx$ 2.17.

\begin{table}
\centering
\begin{tabular}{ |c|c|c|c| } 
\hline
Sample name & Selection criteria & Number & Section ref. \\
\hline
DR12Q & - & 297,301 & \\ 
Hamann+17 & 2.0 $< z_{e} <$ 3.4 & 173,636 & \\ 
& $i$ mag in DR12Q & \\
& well-measured \civ\ & \\ \\ 
Full Sample & \civ\ FWHM & 39,249 & \ref{sec_data_set} \\
& $\ge$2000 \kms\ & \\ 
& well-measured \mgii\ & \\ \\ 
BH Mass & well-measured & 38,966 & \ref{sec:sec_VBHM}, \ref{sec:sec_edd_ratio_dependencies} \\
& \mgii\ wings & & \\ \\ 
Composite & 3.1e45 < $L_{\text{bol}}$ < 2.5e46 & 29,460 & \ref{sec:sec_composite_spec} \\
Sample & \ergs & & \\ \\ 
Color Sample & $r$ \& $z$ mag in DR12Q & 32,365 & \ref{sec:sec_color_dependencies} \\
& 1.8 < $z_e$ < 3.4 & & \\
& $W1$-detected & 25,134 & \\
& cc\_flags = 0000 & & \\ \\ 
High blueshift & > 4,000 \kms\ & 255 & \ref{sec:sec_analysis_individual_spec} \\
\civ\ & > 6,000 \kms\ & 13 & \\
\hline
\end{tabular}
\caption{Selection criteria with our full sample, and total quasar subsample numbers.}
\label{tab:tab_table1}
\end{table}

For more accurate luminosity estimation from BOSS, a flux correction for atmospheric differential refraction was applied from \citet{Margala+16} when available, and otherwise applied a correction from \citet{Harris+16}. We computed monochromatic luminosity ($\lambda \, L_{\lambda}$) using a fitted power law continuum local to \mgii\ near $2800\AA$. This local continuum was extrapolated for the monochromatic flux at $3000\AA$ to compute $L(3000) = 4\pi \, D_{L}^2 \, \, \lambda \, f_{\lambda} $. Where $D_L$ is the luminosity distance, and $f_{\lambda}$ is the flux at wavelength lam. Bolometric luminosity ($L_{\text{bol}}$) was determined using a monochromatic Bolometric correction factor $k_{\text{bol}} \times L(3000) = L_{\text{bol}}$, where $k_{\text{bol}} = 25 \times \left[ L(3000) \, / 10^{42} \,\, \text{erg s}^{-1} \right]^{-0.2} $. This method of correction factor makes considerations for a luminosity-dependent spectral energy distribution over a broad range of black hole masses, accretion rates, and spin, calibrated by \citet{Netzer2019}. We did not do a reddening correction for luminosity, which we note in our analysis in Section \ref{sec:sec_color_dependencies}.

\section{Measurements}
\label{sec_measurements}

In this study we use a dynamic fitting algorithm on every spectra in the \citet{Hamann+17} set to measure line-emission features in the rest frame UV continuum, even for low signal to noise. Specifically, we used the \mgii\ emission-line peaks for systemic redshift, after we applied several Gaussian models for fitting the emission profile, and subtracted a broadened \feii\ template. We assume the \mgii\ doublet emission is one to one ratio, and the emission cloud itself is optically thick ($\tau \gg$ 1). We did not use BOSS catalogue signal to noise values during fitting, because they did not accurately represent the quality of spectral detections near \civ\ and \mgii.

\subsection{Continuum and Line Fits}
\label{sec:sec_cont_line_fit}

The first step is an algorithm that searches the spectrum for narrow spikes above or below the spectra caused by cosmic rays, noise, or other anomalies. These spikes are removed by interpolation from adjacent pixels. We applied a spectral flux correction from \citet{Margala+16} when available, and otherwise applied a correction from \citet{Harris+16}. 

Next we perform an initial preliminary fit to make a rough redshift correction in cases where the BOSS pipeline redshift is too far offset from where we expect \mgii\ emission. We dynamically adjust the size of continuum flux sampled from $\sim$2160 to 3120\AA\ to avoid anomalous features in the spectra when fitting the continuum or emission lines (e.g. fringing as \mgii\ is further redshifted to the end of the spectrograph at larger $z$). We subtracted a modified \feii\ template to remove emission local to \mgii\ \citep[template provided through private correspondence with Yue Shen,][]{Shen+16b}. The template emission is from a theoretical model by \citet{Salviander_2007} scaled to match the empirical template from \citet{Vestergaard_2001} but includes non-zero flux 2780-2830\AA\ under \mgii\ line emission. 

After defining the preliminary continuum, we we mask sharp absorption and make a preliminary single Gaussian fit, for a rough correction to the DR12Q catalogue redshift. We also used the preliminary fit's FWHM as a proxy measurement of \feii\ emission-line broadening. Next we perform the broad absorption checks as before. Again, the \feii\ template is scaled, but before final subtraction the template is broadened according to our preliminary-FWHM of the \mgii\ profile. 

Once we establish a continuum, we check for strong broad absorption with the median fluxes in wavelength intervals 2310-2360\AA\ and 2550-2580\AA\ because broad absorption in these windows by \feii\ is always accompanied by absorption in \mgii. If broad absorption was found, it was flagged for rejection.

Finally, for fitting \mgii\ \lam\lam 2796,2804 line emission we ignore the doublet separation, instead using the mean. We found a robust iterative procedure that begins with a two-component Gaussian model,  which yield good results with a minimum of free parameters \citep[similar to][]{Hamann+17}. If a two-component fit that does not yield physically reasonable profile, or were too similar in shape, we instead fit using a single Gaussian profile. A final peak wavelength measurement is performed with an additional Gaussian fit to the top two thirds of the fitted \mgii\ profile, to mitigate wing asymmetry in the lower one third.

In order to ensure quality spectra measurements we chose a variety of quality standards after fitting, with which to throw out spectra dominated by noise and otherwise ``bad'' fits. We vetted the final sample with the following quality criteria.

A minimum spectral signal-to-noise ratio is chosen as well as our own defined quality standards, such as requiring \civ\ and \mgii\ lines to be well measured based on their fit signal-to-noise >4 in both REW and FWHM. After fitting, we required quality flags for each spectra that indicate no significant problem in the \mgii\ emission-line or continuum fit occurred, and similarly good quality flags from the \civ\ measurements sample. We also placed conservative upper limits on both FWHMs to ensure the algorithm did not incorrectly fit to the continuum instead of the emission-line with our model, and obviously nonphysical. In order to further mitigate incorrect emission-fitting, and allow our general findings in the full sample to be more robust, we have chosen to ignore any spectra containing Broad Absorption Lines (BALs). Spectra were rejected for BALs by the visual inspection flag present in DR12Q, and the Balnicity Index (BI) must be \texttt{bi\_civ\ > 0.0} and \texttt{bi\_civ\ < 1000.0}. Any masking of absorption spikes done to properly fit an emission-line in the \mgii\ emission range must not be a significant fraction of the emission width, and the \mgii\ emission-line peak height must have a good signal to noise. We rejected fits when emission around the fitted \mgii\ profile has a standard deviation >0.75 to further limit the impact of noise when fitting the upper two-thirds of the \mgii\ emission profile. For \feii\ template subtraction quality criteria, we rejected the spectra if the intensity of \feii\ was comparable to the peak intensity of \mgii\ line-emission. This criteria is to limit the bias introduced by the template in our \mgii\ profile fit.

Visual inspections of several thousand of the remaining spectra indicate that the continuum and \mgii\ line fits are generally excellent. We specifically examined results at the extremes of broad and narrow FWHMs, large and small REWs, strong BALs or other broad absorption that might overlap with the \mgii\ emission-line profiles, and spectra with low SNRs in the continuum. The most common problem is fits that underestimate the line peak height and thus overestimate the FWHMs for observed lines that have a strong narrow core on top of much broader wings.

\subsection{Measured Quantities and Blueshifts}
\label{sec:sec_quatities_and_blueshifts}

We measured \civ\ blueshift by comparing the emission redshift of \mgii\ from our fitting to the emission redshift of \civ\ \lam1548,1551 measured by \citet{Hamann+17}. We used the top two thirds fit of \mgii\ emission as the improved systemic redshift. Then we used the \civ\ measurement of the midpoint at half the fitted profile height as the emission velocity shifted from rest \civ. We use negative values to correspond to inflow away from the observer, and positive values as blueshifts, outflowing towards the observer. We visually inspected all $\sim$250 cases with blueshifts >4000 \kms\ to ensure their validity.

For our full sample we obtained measurements for 39,249 quasars, eliminating those with strong BALs and poor data. From these, we measured a multitude of emission line properties which are supplemented by \civ\ measured quantities from \citet{Hamann+17}. For \mgii\ we measured the rest-equivalent width (REW), full width at half maximum (FWHM), second moment line-dispersion ($\sigma$), and centroid redshift. From the continuum fit we took the monochromatic flux ($f_{3000}$) for computing a monochromatic luminosity to approximate Bolometric luminosity. We measured the \mgii\ emission profile with both FWHM and $\sigma$, described in \citet{Peterson+04}.

\subsection{Black Hole Masses}
\label{sec:sec_VBHM}

\begin{figure}
   \includegraphics[width=\columnwidth]{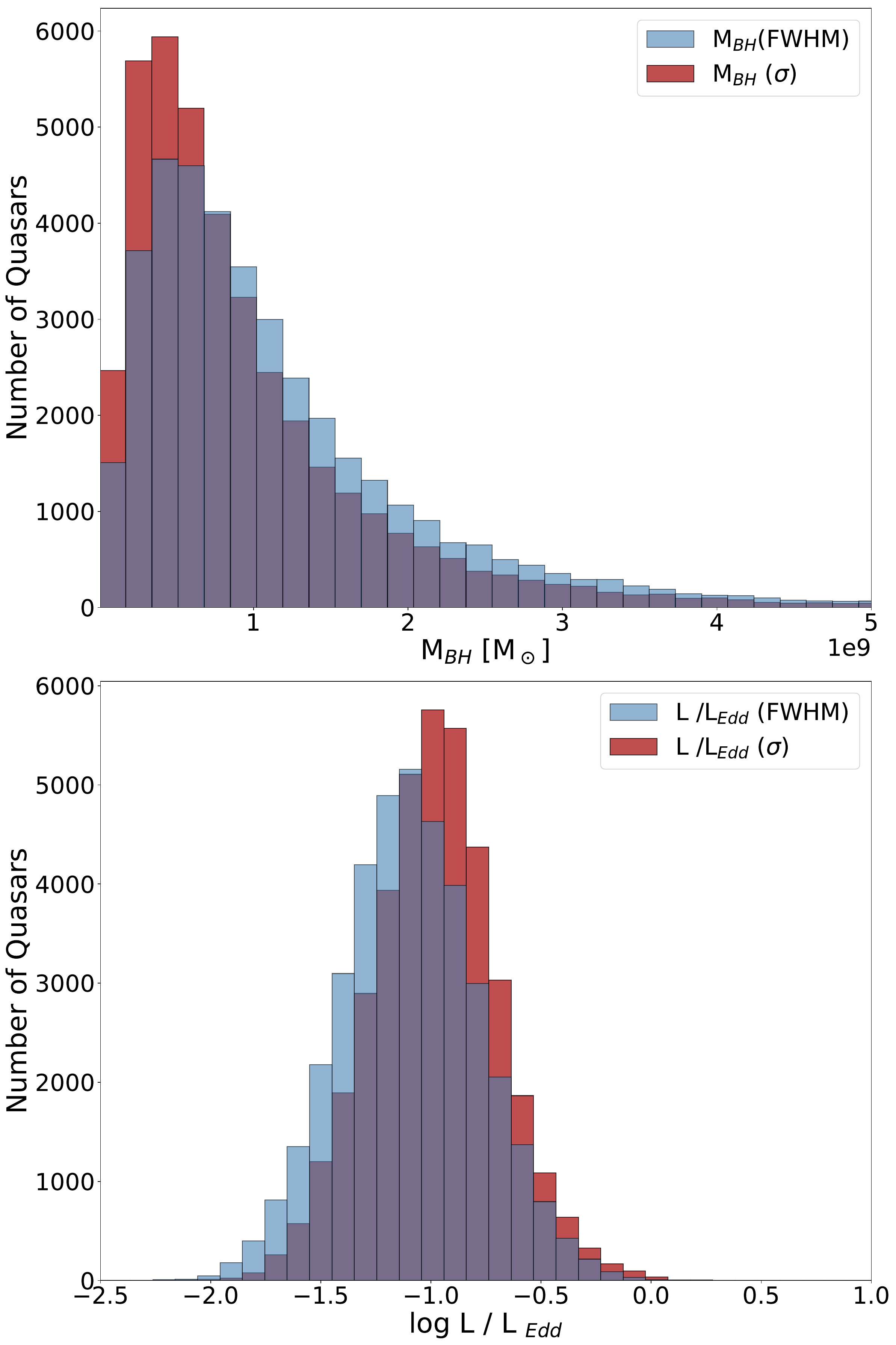}
   \caption{Quasar black hole masses and Eddington ratios for our sample. \textit{Top:} Black hole masses computed from \mgii\ FWHM, or $\sigma$, computed from the fitted emission profile. \textit{Bottom:} Eddington ratios computed from our Bolometric luminosity and black hole mass estimates. To ensure good measurements, we omit linked-Guassian profiles with wing component peak SNR $<$ 0.8, or that have wing FWHM $<$11,000 \kms\ and FWHM $\frac{\text{wing}}{\text{core}}$ ratio $>$ 5.5. Based on our repeated experiments, the M$_{\text{BH}}$ results are not sensitive to these particular values.}
   \label{fig:fig_figure4}
\end{figure}

We computed estimates of quasar black hole mass, assuming that the \mgii\ broad-line region (BLR) has virialized. We consider estimates using both \mgii\ FWHM and $\sigma$, because spectra quality of high-$z$ AGNs can often be insufficient for measuring line dispersion of broad lines \citep{Peterson+04,Shen+08,Wang+09,Shen+12,Woo_2018}. Using the continuum luminosity as a proxy for BLR radius and the broad line width or line dispersion for the virial velocity we estimated virial mass with,

\begin{align*}\label{eq_vbhm}
\centering
\log \left( \frac{ M_\text{BH,vir}}{M_{\odot}} \right) &= \alpha + \beta \, \log \left( \frac{ \text{FWHM}}{10^{3} \text{ km/s } } \right) + \gamma \, \log \left( \frac{ \lambda L_{\lambda}}{ 10^{44} \text{ erg/s} } \right),
\end{align*}

where estimator values $\alpha_\text{FWHM} = 7.02 \pm 0.04$ or $\alpha_\sigma = 7.56 \pm 0.03$, $\beta_{\text{FWHM},\sigma} = 2.00$, and $\gamma_{\text{FWHM},\sigma} = 0.50$ are recommended from \citep[see Table 3 in][]{Woo_2018} for using $L_{3000}$. These values were adopted by calibrated UV mass estimators comparing dispersion of \hb\ to \mgii. There they found continuum luminosity at 3000\AA\ provided better calibration than the line luminosity of \mgii. After making quality cuts based on good \mgii\ signal to noise we have a total of 38,966 quasars (ref. Table \ref{tab:tab_table1}). Comparison distributions for either method are shown in Figure \ref{fig:fig_figure4} and Figure \ref{fig:fig_figure11}. FWHM tends to overestimate large widths, and underestimate the narrow widths. Most of our fits use single Gaussians, and therefore driven by FWHM. In this work, we present results for both FWHM and $\sigma$, but opt to use FWHM to parameterize line width. \\

\section{Results \& Analysis}\label{results}

\begin{figure}
	\includegraphics[width=\columnwidth]{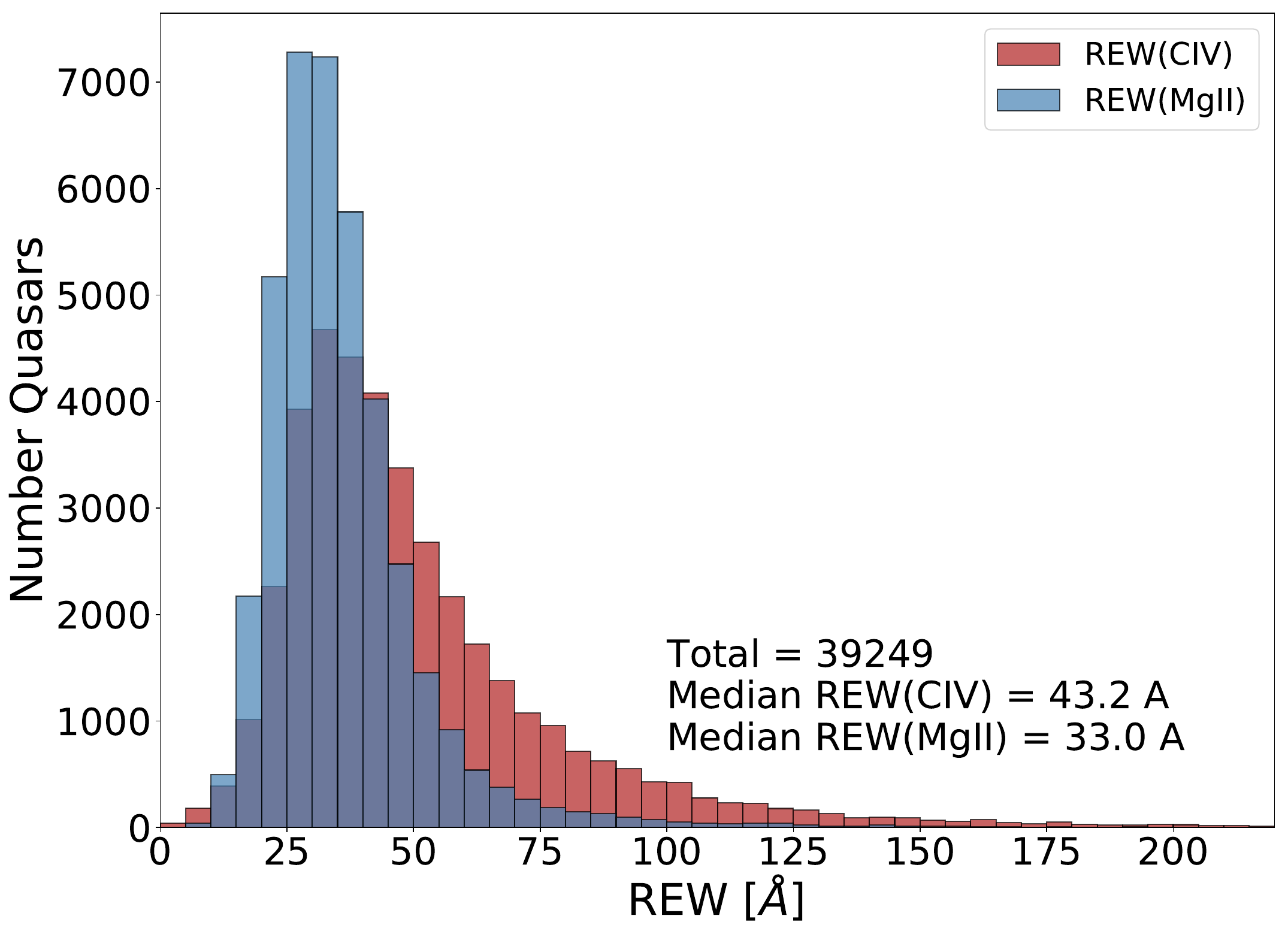}
    \includegraphics[width=\columnwidth]{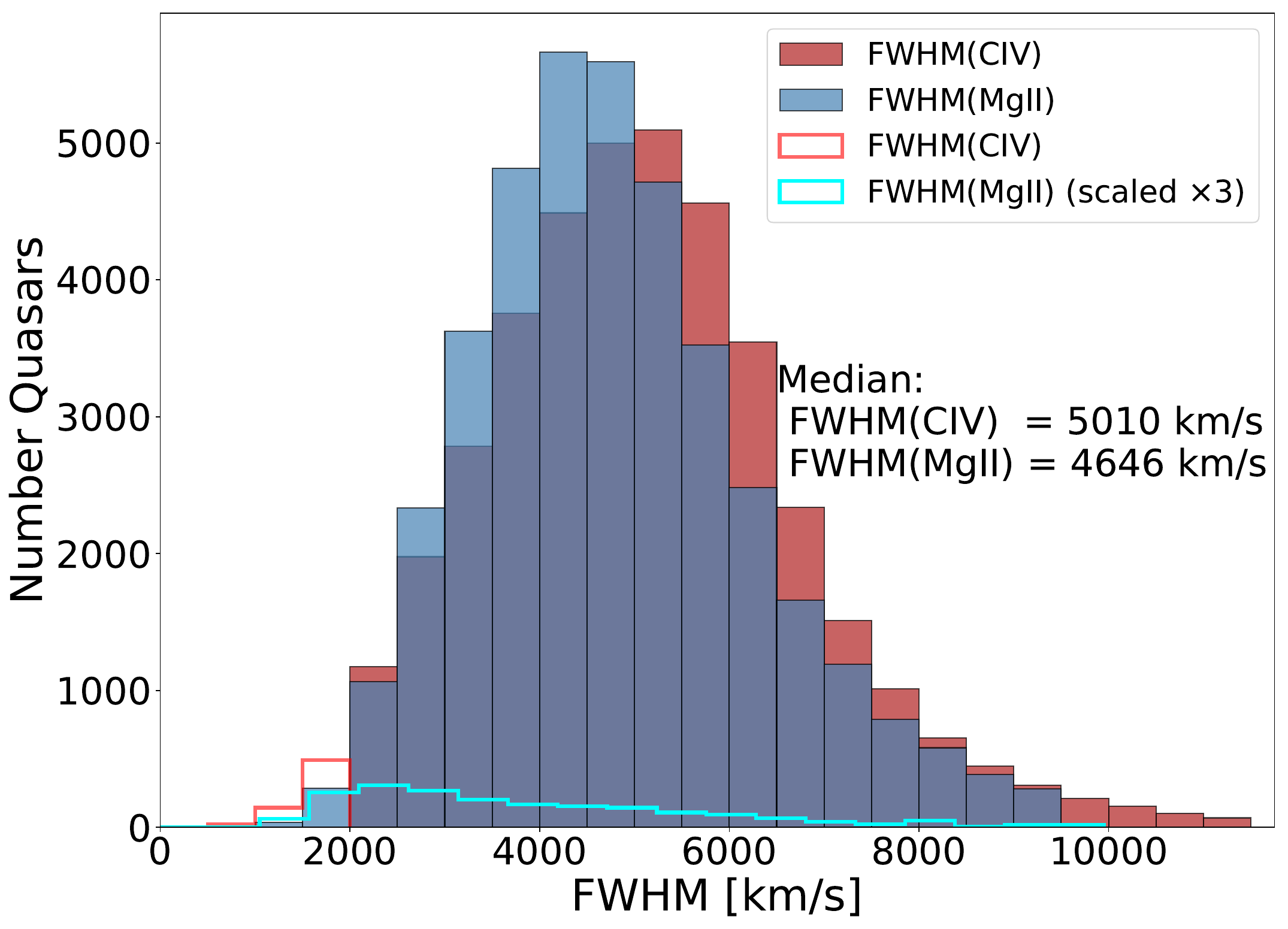}
    \includegraphics[width=\columnwidth]{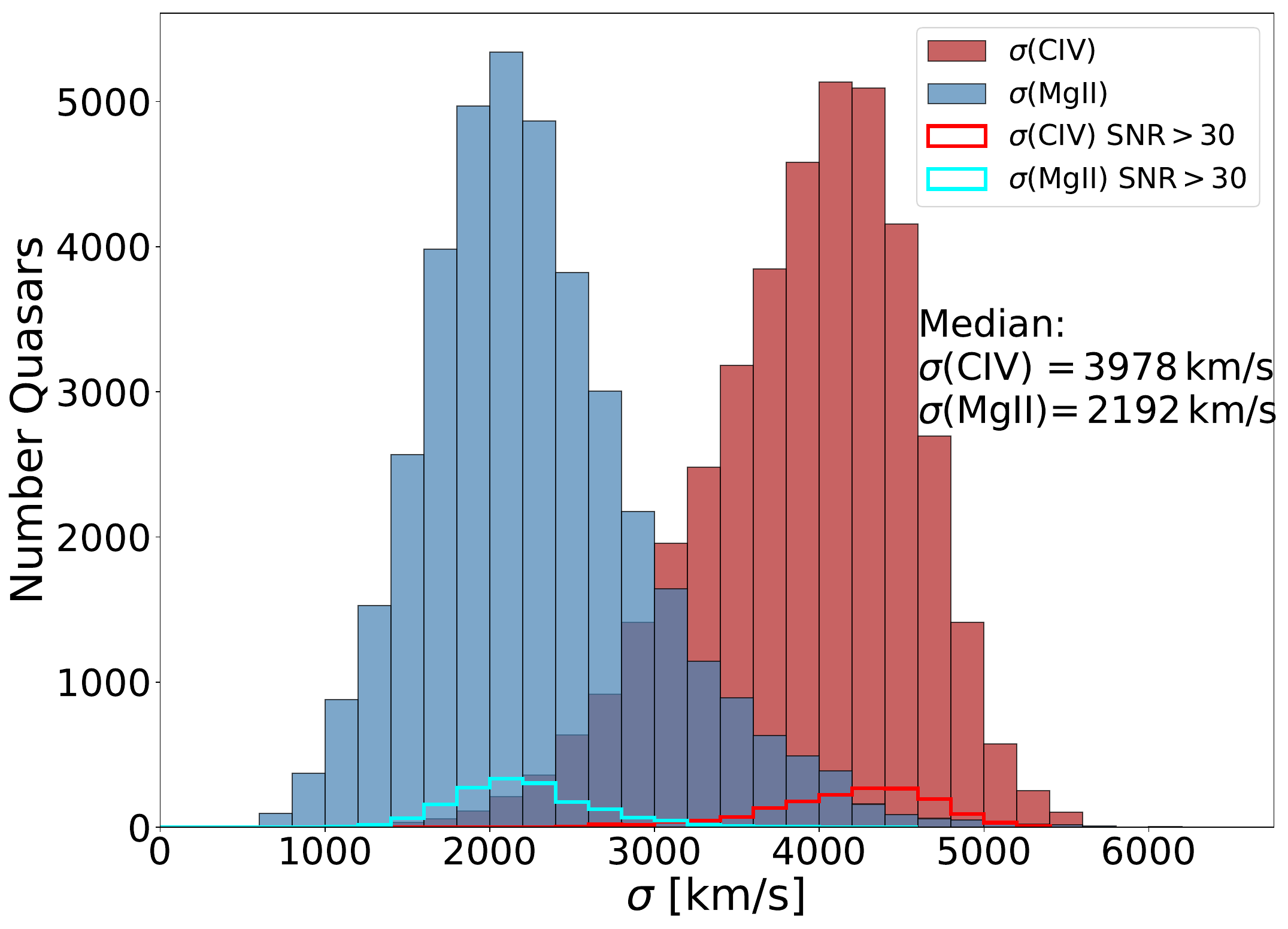}
    \caption{Distributions of measured quantities for our sample. \textit{Top:} REW distributions the full sample of quasars for both \civ\ (blue histogram) and \mgii\ (red histogram) emission lines. \textit{Middle:} FWHM distributions for all quasars for both \civ\ (red histograms) and \mgii\ (blue histograms) emission lines. Distinction between broad and narrow line quasars is made by \civ\ emission FWHM, where broad emission are defined by \civ\ FWHM $\ge$2000 \kms\  (solid red histogram), and narrow as \civ\ FWHM $<$2000 \kms\ (outlined red histogram). The solid blue histogram shows \mgii\ FWHM for quasars with \civ\ FWHM $\ge$ 2000 \kms, and the outlined cyan histogram shows \mgii\ FWHM for quasars with \civ\ FWHM < 2000 \kms. \textit{Bottom:} $\sigma$ distribution for single and linked-Gaussian profiles used in our  full sample.}
    \label{fig:fig_figure2}
\end{figure}

\begin{figure}
	\includegraphics[width=\columnwidth]{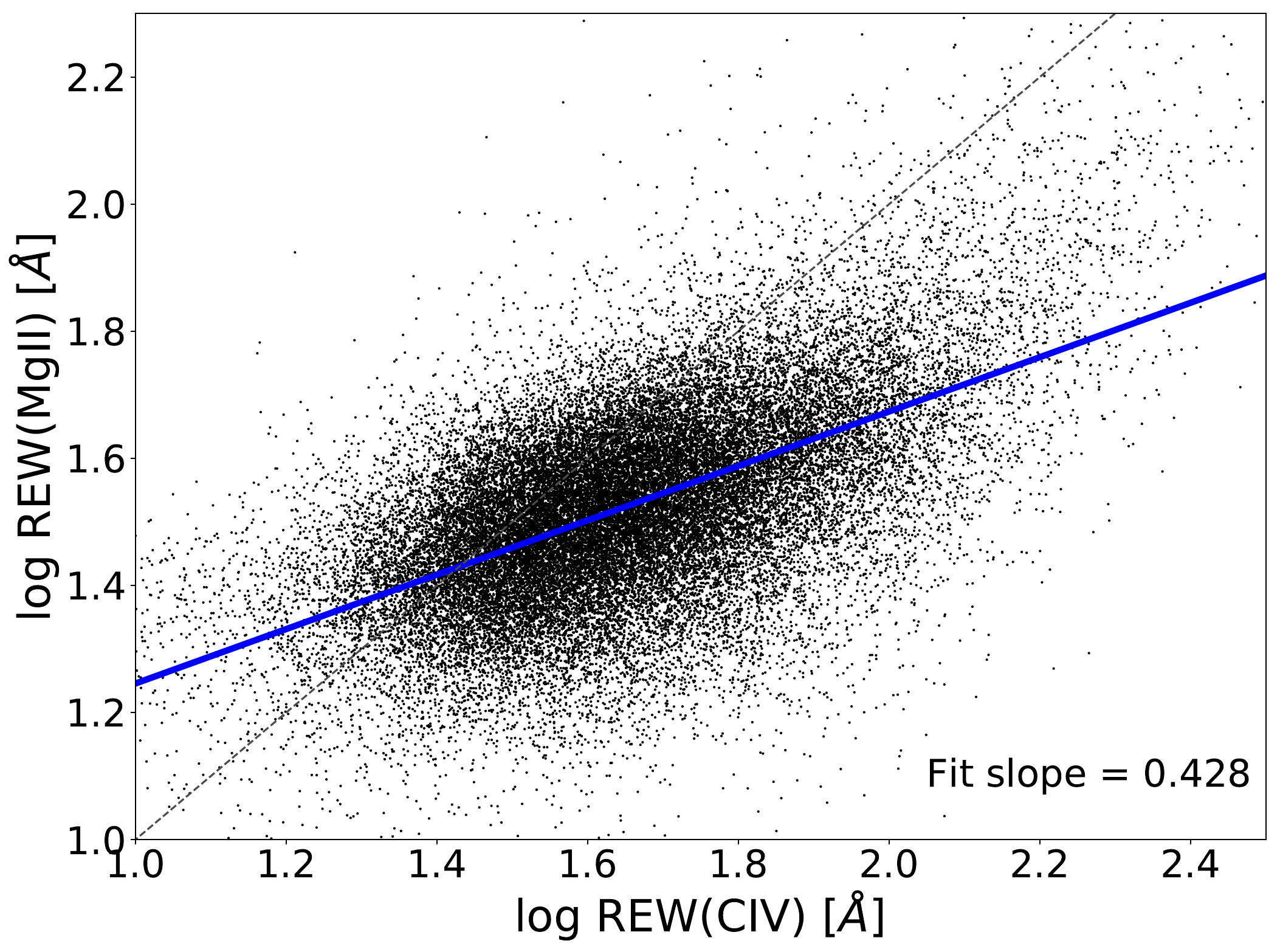}
    \includegraphics[width=\columnwidth]{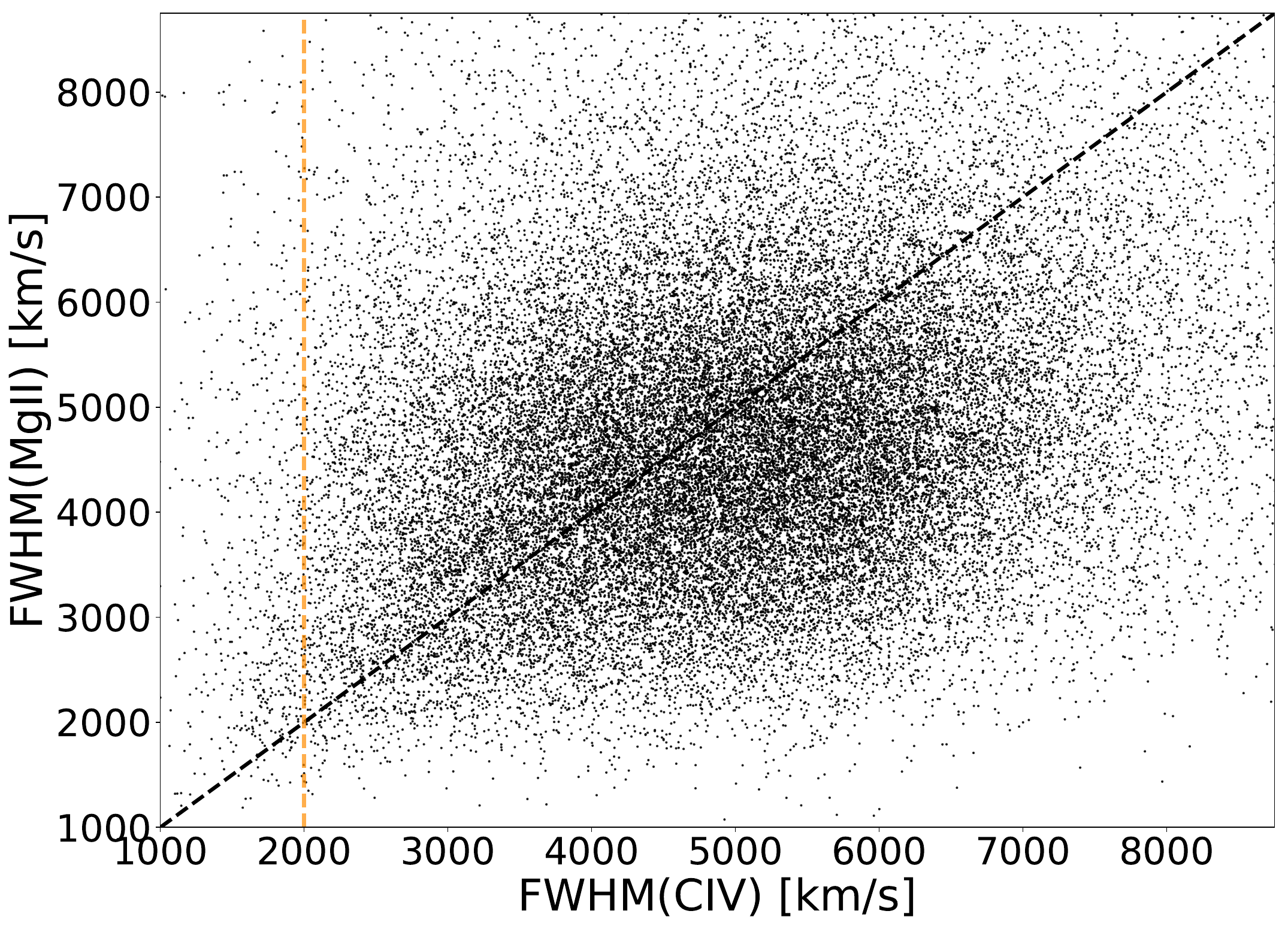}
    \includegraphics[width=\columnwidth]{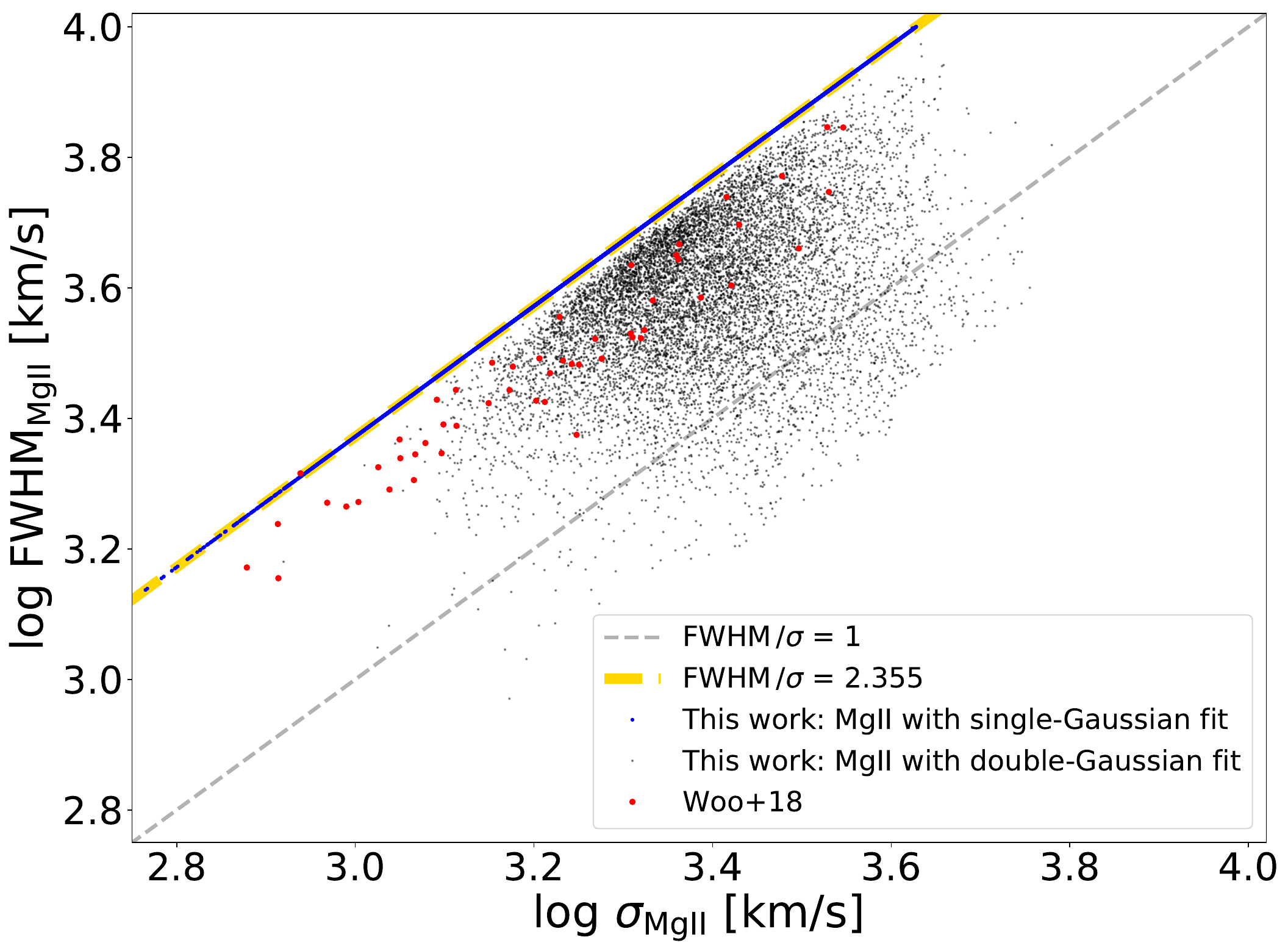}
    \caption{Measured emission line property comparisons. \textit{Top:} REW of \civ\ vs \mgii\ for the full sample of quasars. \textit{Middle:} The ratio of FWHM \civ\ vs \mgii\ for the full sample. The yellow dashed line is our cutoff for broad \civ\ emission quasars. \textit{Bottom:} Log FWHM vs log $\sigma$ of \mgii, for all broad \civ\ emission quasars in the BH Mass sample (see Table \ref{tab:tab_table1}). Black points are our linked-Gaussian profile fits, blue points are from single-Gaussian fits, and the yellow dashed line highlights the relationship of FWHM to $\sigma$ for a Gaussian emission-profile. About 75 percent of the full sample have \mgii\ measured with a single-Gaussian, yielding FWHM$\,$/$\sigma$ = 2.355. There is some overlap between our sample and that of \citet{Woo_2018}, despite their sample redshifts $z$ $\approx$ 0.3$-$0.6. All panels show one to one as a grey dashed line.}
    \label{fig:fig_figure3}
\end{figure}

For a basis in presenting numbers of quasars, etc., we present distributions of our measured parameters. Figures \ref{fig:fig_figure2} \& \ref{fig:fig_figure3} present distributions in redshift, Bolometric Luminosity ($L_{\text{bol}}$), REW, FWHM, and $\sigma$ for both \civ\ and \mgii.

Figure \ref{fig:fig_figure2} shows the distributions of line strengths and widths. \civ\ has slightly larger REW than \mgii, but have similar distributions. In the general population with broad \civ\ emission FWHM has similar distributions for \civ\ and \mgii. There are many examples of quasars with FWHM of \civ\ and \mgii\ both being large or small but there may be extreme cases where differences in width are exacerbated, and may correlate with other quasar characteristics. The last panel shows our integrated line dispersion for \civ\ and \mgii. The distributions appear similar, but there is a clear offset between the two emission lines.

Figure \ref{fig:fig_figure3} displays distributions of line strength and width of \civ\ and \mgii. Emission line strengths generally appear to correlate in REW of \civ\ and \mgii, but flatter than one to one. However, spectra showing strong \civ\ emission generally have weaker \mgii\ by at least 0.2dex. Emission line widths between \civ\ and \mgii\ FWHM are generally uncorrelated. Plotting \mgii\ FWHM against $\sigma$, we impose additional quality cuts to fits with linked-gaussian fits to omit large wing profiles of low signal to noise ratio (see Table \ref{tab:tab_table1}). We compare the general distribution to robustly measured quasars at lower redshifts by \citet{Woo_2018}, and show overlap in line widths. We can see in the distribution of our sample that profiles fit with a double-Gaussian (25 percent) extend to larger values of $\sigma$ than those fit with a single-Gaussian.

Our broad goal is to understand what physical or evolutionary properties of quasars lead to large BEL blueshifts, and fast powerful outflows capable of feedback. We first quantify the \civ\ blueshifts, and then investigate other quasar properties and physical conditions for empirical relationships with blueshift.

\subsection{Blueshifts Across the Sample }\label{sec:sec_blueshift_distributions}

\begin{figure}
	\includegraphics[width=\columnwidth]{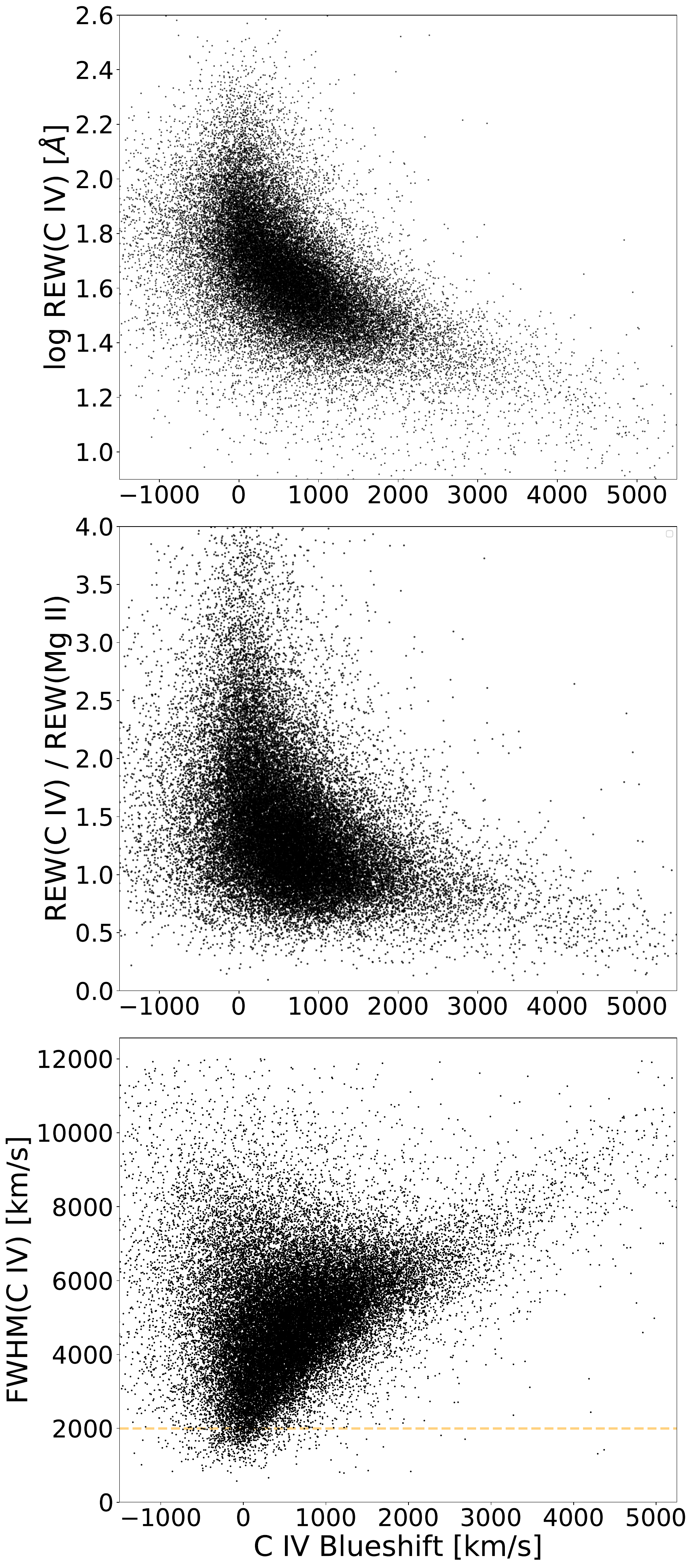}
    \caption{Measured quantities from our sample where all distributions exclude FWHM(\civ) <2000 \kms, unless otherwise stated. \textit{Top:} REW(\civ) vs \civ\ velocity shift, exhibiting the characteristic distribution that has emerged in the literature as quasar redshift measurements have been constrained. \textit{Middle:} The ratio of REW(\civ) to REW(\mgii) vs \civ\ velocity shift, showing that for higher blueshifts the REW(\civ)\,/REW(\mgii) $\le \sim$1. \textit{Bottom:} FWHM(\civ) vs \civ\ velocity shift of all quasars in our full sample, with a yellow dashed line indicating quasar cutoff for broad \civ\ line quasars for our sample. Most quasars appear to follow a trend for broader \civ\ emission profiles with higher \civ\ blueshift.}
    \label{fig:fig_figure5}
\end{figure}

Figure \ref{fig:fig_figure5} shows distributions of \mgii\ REW, \civ\ REW and FWHM vs blueshift. There are several quasars that appear to show redshifted \civ. These redshifts may be due to the uncertainties and relative motions of \civ\ and \mgii, and not high velocity inflow. We visually inspected all 255 cases with measured blueshifts greater than 4000 \kms. We were concerned that these extreme cases have broad and weak \civ\ lines (see Section \ref{sec:sec_discuss_individual_spectra} below) that might be poorly or incorrectly measured. However, the vast majority of these high blueshift cases were found to have good fits. Most of the sample resides between blueshifts of 0 and 1500 \kms, with a median \civ\ blueshift of about 530 \kms. We measure reliable blueshifts for 1178 quasars with velocities >2500 \kms. From those, 255 quasars with blueshifts above 4000 \kms, 13 above 6000 \kms, and the highest well-measured \civ\ velocity $\sim$~7000 \kms. Figures of \civ\ REW vs its blueshift has become common in the literature \citep[e.g.,][]{Richards+11,Coatman+16,Coatman+19,Rankine+20,Temple+20,Temple+23}. This work differs in that we exclusively use the \mgii\ emission line as the proxy measurement for systemic redshift, where others use catalogue redshifts from template fitting or a combination of low-ionization lines, which may include \mgii. Our distribution is in broad agreement with other studies, and have comparable sample sizes, but we extend to further extremes in the \civ\ parameter space of line strength/weakness and blueshift. The top two panels of Figure \ref{fig:fig_figure5} show not only that \civ\ REWs are small at large blueshifts, but they are small relative to the low-ion line \mgii. The bottom panel shows that \civ\ lines can be broad at any velocity shift, but at large blueshifts they are almost exclusively broad. 

\begin{figure}
	\includegraphics[width=\columnwidth]{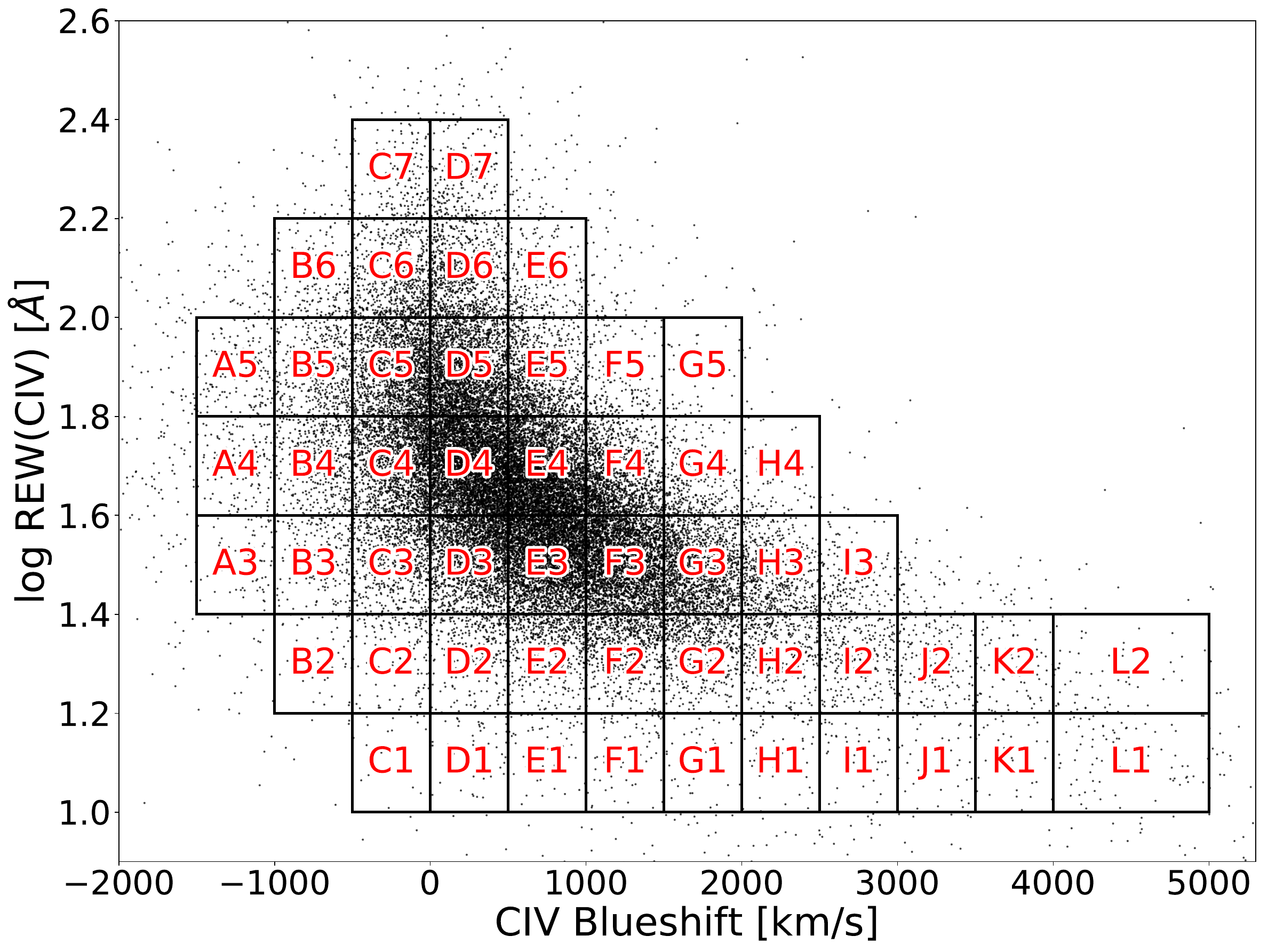}
    \caption{REW(\civ) vs \civ\ velocity shift, identical to the top panel of Figure \ref{fig:fig_figure5}, but with partitions labelled for further analysis (see Sections \ref{sec:sec_composite_spec}, \ref{sec:sec_edd_ratio_dependencies}, and \ref{sec:sec_color_dependencies}. We choose box sizes which can resolve overall trends, but not so few spectra to have poor statistics for each box. Cells are 0.2dex in height and and 500 \kms\ in width, except for column L, which needed to be 1000 \kms\ in width to include a sufficient number of quasars. Every partitioned cell contains at minimum $\sim$100 objects.}
    \label{fig:fig_figure6}
\end{figure}

\begin{figure*}
	\includegraphics[width=\textwidth]{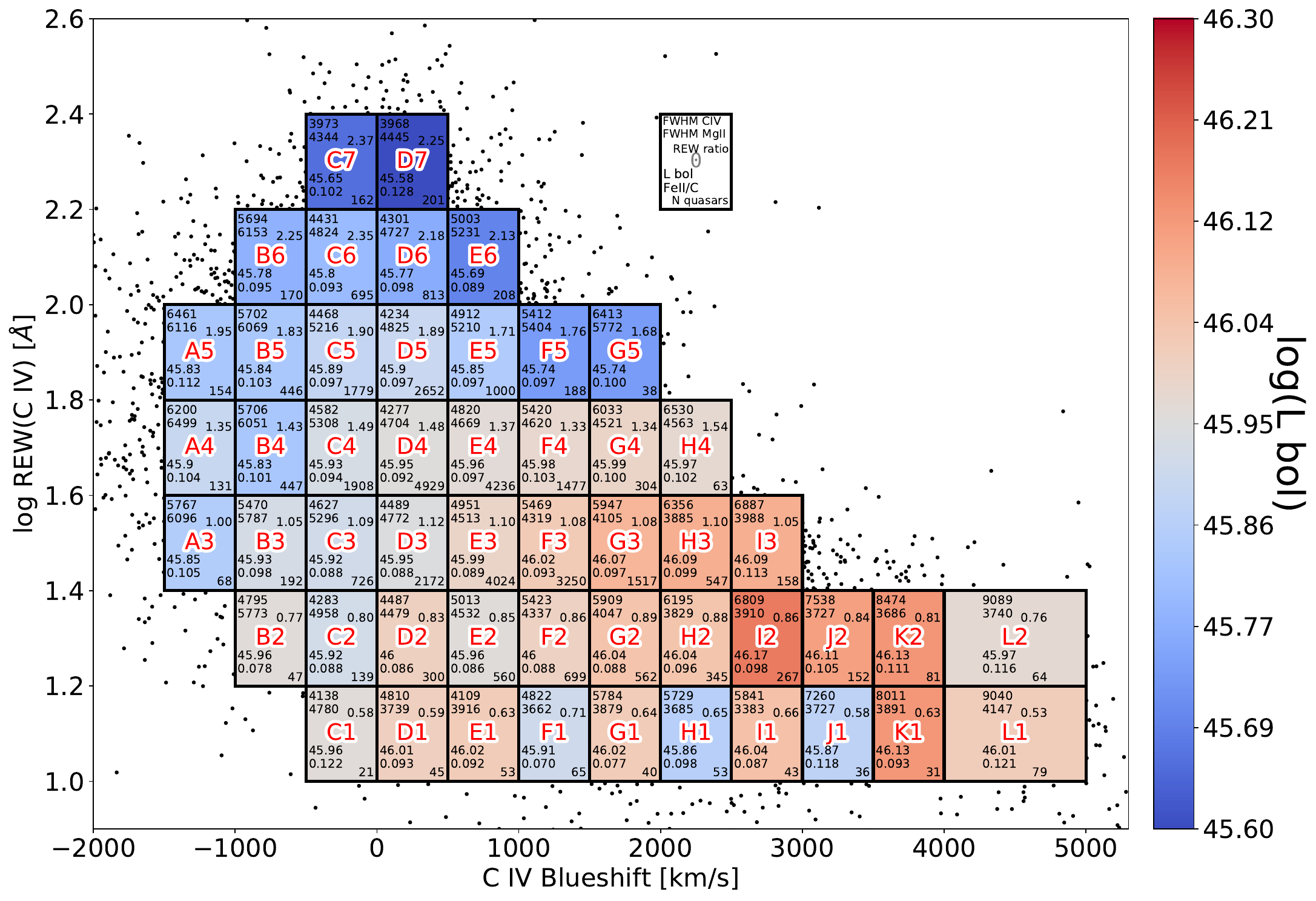}
    \caption{REW(\civ) vs \civ\ velocity shift of our sample with regions color coded for median $L_\text{bol}$. Within each boxed region we present median values from the quasar spectra. The white box at the top right gives a legend to where the values are located. These values include, FWHM of \civ\ and \mgii\, REW ratio of \civ\ to \mgii\, $L_{\text{bol}}$, the relative strength of \feii\ to continuum, and the number of quasars in the boxed region. The color of the region indicates the median 
    \imw\ color within the box. }
    \label{fig:fig_figure7}
\end{figure*}

Figure \ref{fig:fig_figure6} is identical to Figure \ref{fig:fig_figure5}, with subdivisions in to boxes to further inspect local median quantities across the parameter space. This partitioning allows us to vary \civ\ REW and blueshift independently, and show how medians of other measured quasar properties correlate. Figure \ref{fig:fig_figure7} displays median values of Bolometric luminosity, and  several median properties for quasars in each box. This figure more quantitatively shows the blueshift trends with \civ\ line strength and width. The vertical trend of larger \civ\ REW with smaller $L_{\text{bol}}$ is the Baldwin Effect \citep{Baldwin77}, but there also appears to be a weak trend for larger $L_{\text{bol}}$ with larger blueshifts (see Section \ref{sec:sec_edd_ratio_dependencies}). Median \civ\ FWHM of each box is uniform in weak blueshifts, and dramatically increases for large blueshift quasars. At high blueshifts (>3000 \kms), \civ\ rarely exhibits line strength REW above 40 \AA. In this high blueshift regime \civ\ becomes almost always broad, with median FWHM > 7000 \kms, and is broad relative to \mgii\ (median FWHM about 3700 \kms). Median Eddington ratio in each box also appears to increase with \civ\ blueshift. Median strengths of \feii\ relative to the continuum, which is used to subtract the \feii\ complex, appear stronger for large blueshift.

We further inspect other color-coded versions of this figure, and discuss extinction effects on luminosity due to quasar reddening, in Section \ref{sec:sec_color_dependencies}. We inspect median composite spectre to confirm trends in median quantities with blueshift in Section \ref{sec:sec_composite_spec}.

\begin{figure}
    \includegraphics[width=\columnwidth]{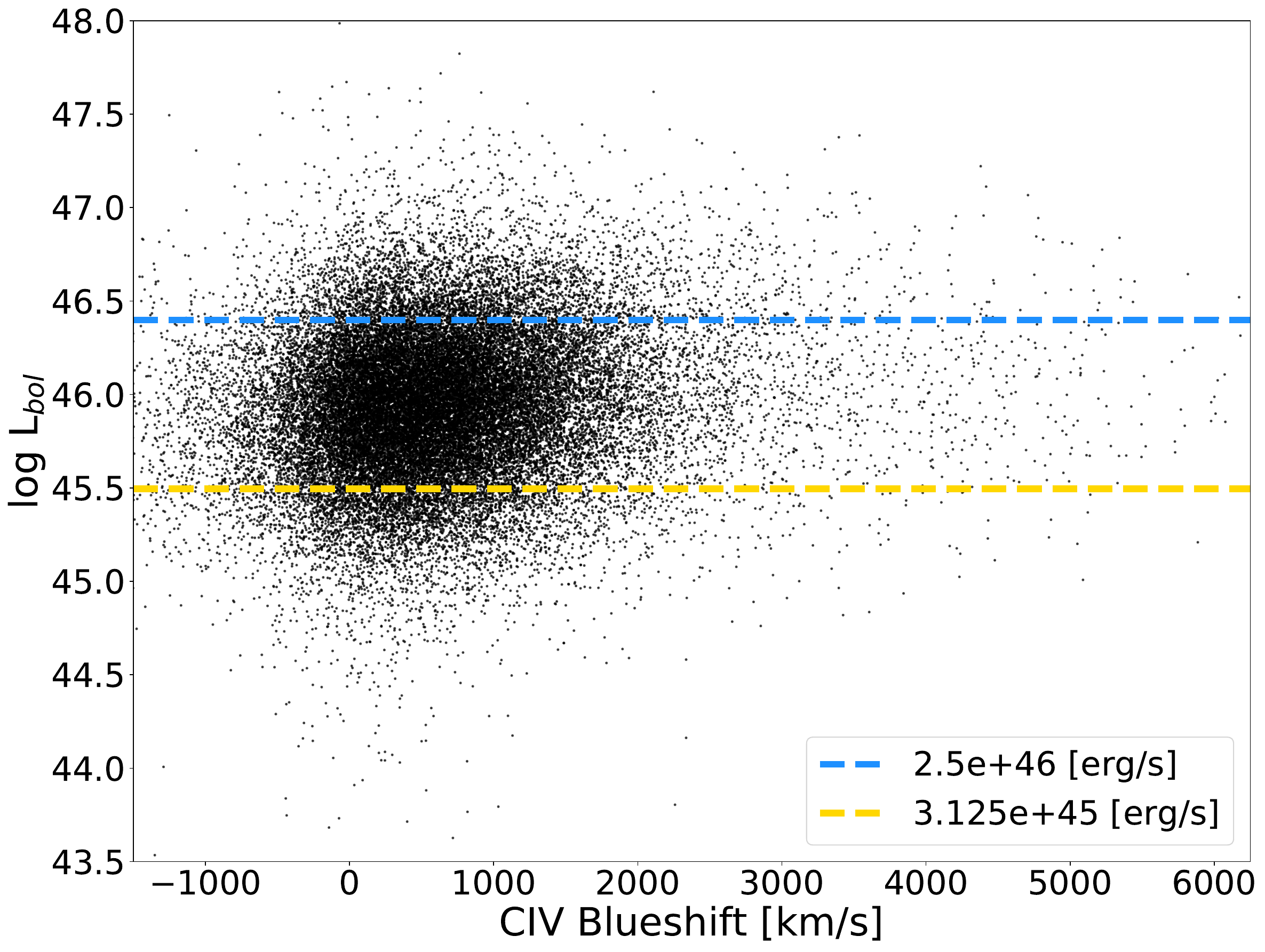}
    \caption{Bolometric luminosity vs blueshift, with boundaries defined in luminosity to isolate trends in blueshift without biasing composite spectra to have both high blueshift and high luminosity.}
    \label{fig:fig_figure8}
\end{figure}

\subsection{Composite Spectra}
\label{sec:sec_composite_spec}

Here we generate composite spectra to see overall trends in quasar properties with varying \civ\ line strength and blueshift. We use medians of selected rows and columns across our partitioned boxes in Figure \ref{fig:fig_figure6}, to show variation of spectral features across a wide range of \civ\ REW and blueshifts. Implications of found correlations will be discussed in Section \ref{sec:sec_discussion}.

We isolate trends with respect to \civ\ blueshift from luminosity by constraining $L_{\text{bol}}$, by setting an upper and lower bound for inclusion, when generating medians. Figure \ref{fig:fig_figure8} shows the distribution of computed $L_{\text{bol}}$ across \civ\ blueshifts for our sample. We construct these median composite spectra from quasars within the luminosity range 3$\times$10$^{45}$ \ergs\ < $L_{\text{bol}}$ < 2.5$\times$10$^{46}$ \ergs. Imposing this constraint ensures that median luminosities of our composite quasars remain nearly constant across the full range of \civ\ blueshifts, and still retains a large sample of $\sim$30,000 quasars (see Table \ref{tab:tab_table1}).

Table \ref{tab:tab_table2} presents quasar parameter medians from the Composite Sample in Table \ref{tab:tab_table1}, grouped by \civ\ velocity from $-$1500 to 6000 \kms, and in bins of 500 \kms. We chose this bin size to include a significant, minimum number of quasars at large blueshift bins ($\sim$100 quasars), without obscuring trends by broadening the composite emission profile. We made luminosity cuts to isolate trends that are independent of $L_{\text{bol}}$ and the Baldwin Effect (see Table \ref{tab:tab_table1} and Figure \ref{fig:fig_figure8}). Our median $\lambda L_\lambda$ luminosities are are roughly constant across the wide range in velocity (within about 12 per cent of the overall median), confirming our luminosity criteria is effective, and shows that any other trends seen are not explained by significant variations in luminosity.

Figure \ref{fig:fig_figure9} shows median spectra from the Composite Sample in Table \ref{tab:tab_table1}, from $-$1500 to 6000 \kms, and associated with the medians in Table \ref{tab:tab_table2}. Most notable, is \civ\ FWHM and REW correlation with blueshift. The line strength of \civ\ weakens, indicated by the median REW decreasing from $\sim$50 to 10 \AA, as blueshift increases. Conversely, the line broadens with increasing blueshift. From median \civ\ velocity $\sim$260 to 5260 \kms, the FWHM more than doubles from $\sim$4300 to 9400 \kms. \mgii\ profile width as measured by FWHM, or $\sigma$, does not display dramatic broadening or narrowing with large changes in \civ\ blueshift. 

Although we describe the blueshifts in terms of a parameter similar to the line centroid \citep[Section 5.8 in][]{Hamann+17}, the blue wings in the highest blueshift cases extend beyond 11,000~\kms. In addition, the changing appearance of the \civ\ line profiles as we look from low to high blueshifts suggests there is a decrease in the emission from material at low velocities, rather an increase in emission from high-velocity gas. 

Overall the spectra show changes in many line strengths, including \heii, with increasing \civ\ blueshift. Near to \civ, we can see \heii\ also gradually weakening until there is no line emission when \civ\ is at the its most extreme blueshift. It is important to note that while \heii\ and \civ\ are getting weaker at large blueshifts, other lines from lower-ion species (\cii], \mgii, etc.) stay the same or get slightly stronger. Some lines become weaker relative to others, such as weaker \ciii] \lam1909 relative to \aliii\ and \siiii]. Some low ionization lines appear stronger, such as \cii] and Fe complexes, with increasing \civ\ blueshift. REW for \mgii\ only varies slightly in comparison to \civ\ line strength across blueshift velocities, which is also visible in other low-ionization line strengths. 

Low-ionization lines, such as \oi\ and \cii, do not shift in velocity relative to \mgii, and appear to agree with the \mgii\ emission centroid for systemic redshift. Any velocity shifts in these lines relative to a true quasar systemic would not be evident in our data.

Finally, \civ's asymmetric broadening and weakening is also present in high-ionization lines, such as \siiv\ and \nv, but less apparent. However, the magnitude of all of these changes is less than what appears in \civ, consistent with less participation in a BLR outflow by lower-ionization lines.

Figure \ref{fig:fig_figure10}'s top two panels show the median composite spectra from a single column from Figure \ref{fig:fig_figure7}, at \civ\ blueshifts 0 to 500 \kms and REW 16 to 251 \AA. Both emission profiles of \civ\ and \mgii\ remain symmetric and gradually weaken along with \civ. All other lines in the spectrum likewise decrease in strength. Low-ionization lines, and Fe complex, also decrease in strength with \civ\ REW, which is not the case when \civ\ is weak and strongly blueshifted.

Figure \ref{fig:fig_figure10}'s bottom two panels display the same spectral range, but using composites from specific partitioned boxes in Figure \ref{fig:fig_figure6}, following diagonally along the bulk of the distribution. This composite shows the same trends seen in Figure \ref{fig:fig_figure9} more cleanly. Again, we see the low ionization lines remaining strong (e.g., \mgii, \cii, and Fe complex), while other lines weaken with \civ. Some high ionization lines, like \siiv\ and \nv, blueshift in a similar asymmetric pattern as \civ's profile.

\subsection{Eddington Ratio Dependencies}
\label{sec:sec_edd_ratio_dependencies}

Eddington ratio shows a nearly uniform correlation with increasing \civ\ blueshift.
Table \ref{tab:tab_table2} shows Eddington ratio in the Composite Sample increases from 0.08 to 0.14, with larger \civ\ blueshift. 

Figure \ref{fig:fig_figure11} shows the distribution of Eddington ratios against blueshift for our BH Mass sample in Table \ref{tab:tab_table1}, with black hole mass computed using both FWHM and dispersion methods described in Section \ref{sec:sec_VBHM}. Using either \mgii\ FWHM or $\sigma_\text{\mgii}$ to compute \ledd\ yields a similar distribution. We note that the $\sigma_\text{\mgii}$ has less scatter than the FWHM method. However, most \mgii\ profiles were fit using a single Gaussian, which conversion from FWHM to $\sigma$ is a constant, and so we choose to use FWHM for our distribution comparisons. Figure \ref{fig:fig_figure11} illustrates that the method used to measure black hole mass does not have significant effect on Eddington ratio trends.

Figure \ref{fig:fig_figure12}'s top panel shows log Eddington ratio color-coded across the distribution of \civ\ REW versus blueshift. The trend in Eddington Ratio is stronger, and more coherent, than with Bolometric luminosity seen in Figure \ref{fig:fig_figure7}. It is evident that the trend in Eddington ratio is not a simple correlation with \civ\ blueshift, and that REW is an important third parameter, because the trend does not strictly follow  left to right across the REW and blueshift distribution.

\subsection{Color Dependencies \& Reddening Corrections}
\label{sec:sec_color_dependencies}

We show distributions of the Color Sample in Table \ref{tab:tab_table1}, requiring quasar colors using \textit{r} and \textit{z} filters from the SDSS (\citealt{York2000}; \citealt{Alam2015}), and/or $W1$ band from WISE (\citealt{Wright_2010}, \citealt{Yan_2013}), as provided in the BOSS DR12Q quasar catalogue (see \citealt{Paris2014} and \citealt{Paris2017}). \textit{r}, \textit{z}, and $W1$ filters fall approximately in the rest wavelengths 1960 \AA, 2880 \AA, and 1 \mum\ at our Full Sample median redshift. These colors were chosen to avoid contamination by the wings of the \civ\ profile (specifically in \textit{r}), especially if it is blueshifted, but still a wide enough sampling of emission determine a slope of the UV continuum.

Figure \ref{fig:fig_figure12}'s bottom two panels show the Color Sample with \civ\ REW versus blueshift, except color coded to \rmwo\ and \rmz. These quasar color panels show the same trend as Eddington ratio, although $W1$ limits the sample more due to lower sensitivity. These panels show that the color trend is not simply 2-dimentional with color vs blueshift, seen in Figure \ref{fig:fig_figure13}. Figure \ref{fig:fig_figure13} presents the distribution of \rmz\ and \rmwo\ colors compared to blueshift. \rmz\ shows a distribution similar to figures including \civ\ REW, with \rmz\ decreasing with increasing blueshift. This trend is less strong in the \rmwo\ distribution.

Quasars with smaller $L_{\text{bol}}$ and Eddington ratios are also redder. A reddening correction could therefore weaken trends we find with $L_{\text{bol}}$ and Eddington ratio. However, it’s not clear that a correction is appropriate given that there might be real difference in the SED unrelated to dust reddening. If we assume these quasars have the same intrinsic SED, reddening corrections indicated by the color differences would be quite small. We confirm this within the range of \rmz\ values from Figure \ref{fig:fig_figure12}, and the median redshift. Using a typical dust reddening formula \citep[similar to those used in][]{Richards03}, we get E(B-V) = 0.105 and extinction A$_{3000}$ = 0.57 magnitudes, or a factor of 1.69 flux suppression.

\subsection{Individual Spectra}
\label{sec:sec_analysis_individual_spec}

\begin{figure*}
 	\centering
    \includegraphics[width=0.94\textwidth]{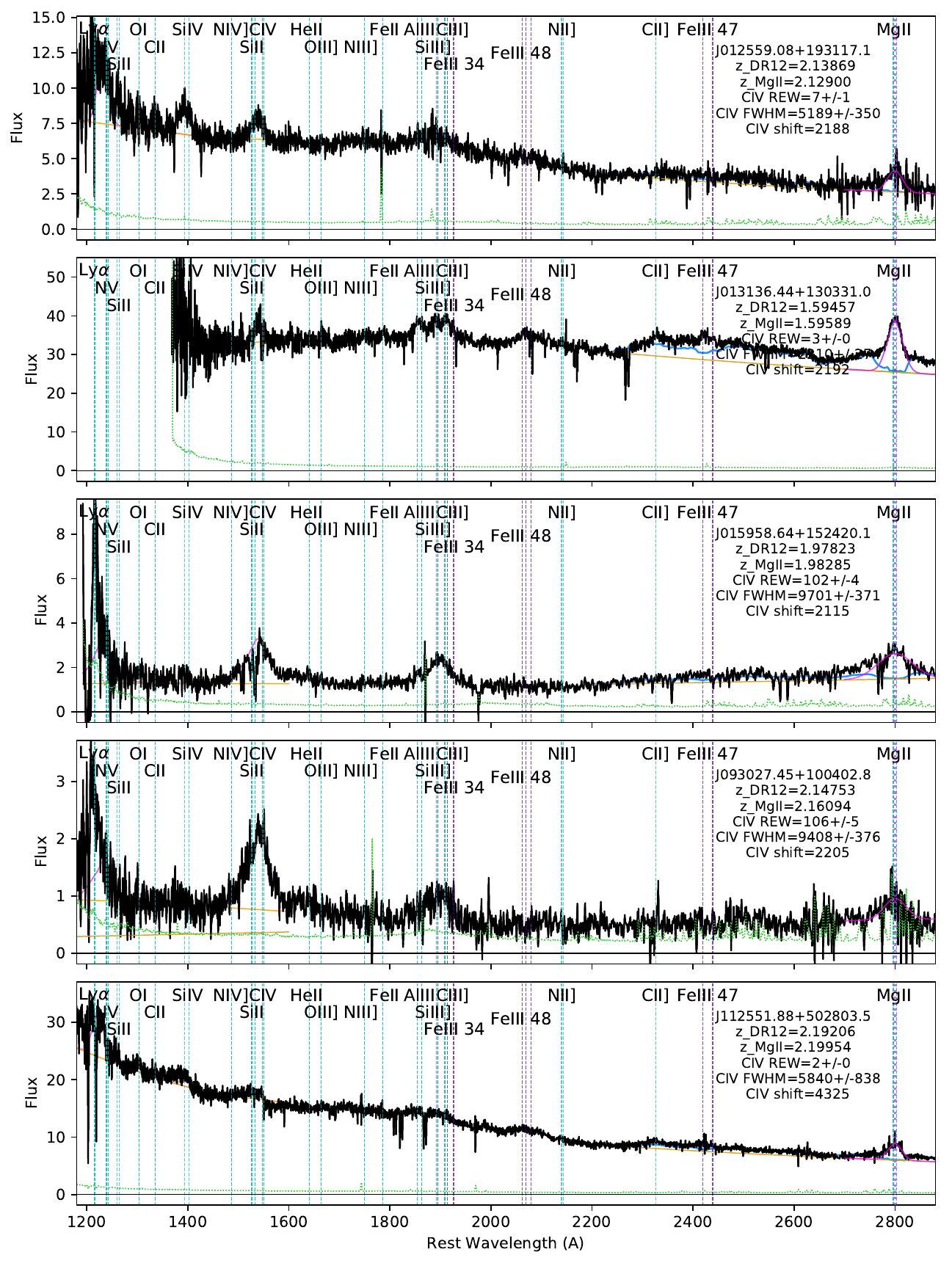}
    \caption{Spectra with varied \civ\ REW, and large blueshifts (>2000 \kms). Black is spectral data, green is the error spectrum, the orange curves are fitted power laws to the local continuum of \civ\ or \mgii, and dark blue curve is the broadened \feii\ template near \mgii. Magenta are the full \civ\ and \nv\ profile fits \citep[from][]{Hamann+17}, or the \mgii\ profile fits from this work. Cyan curves are the two Gaussian components used when necessary to fit the \mgii\ profile.}
    \label{fig:fig_figure14}
\end{figure*}

\begin{figure*}
 	\centering
 	\includegraphics[width=0.95\textwidth]{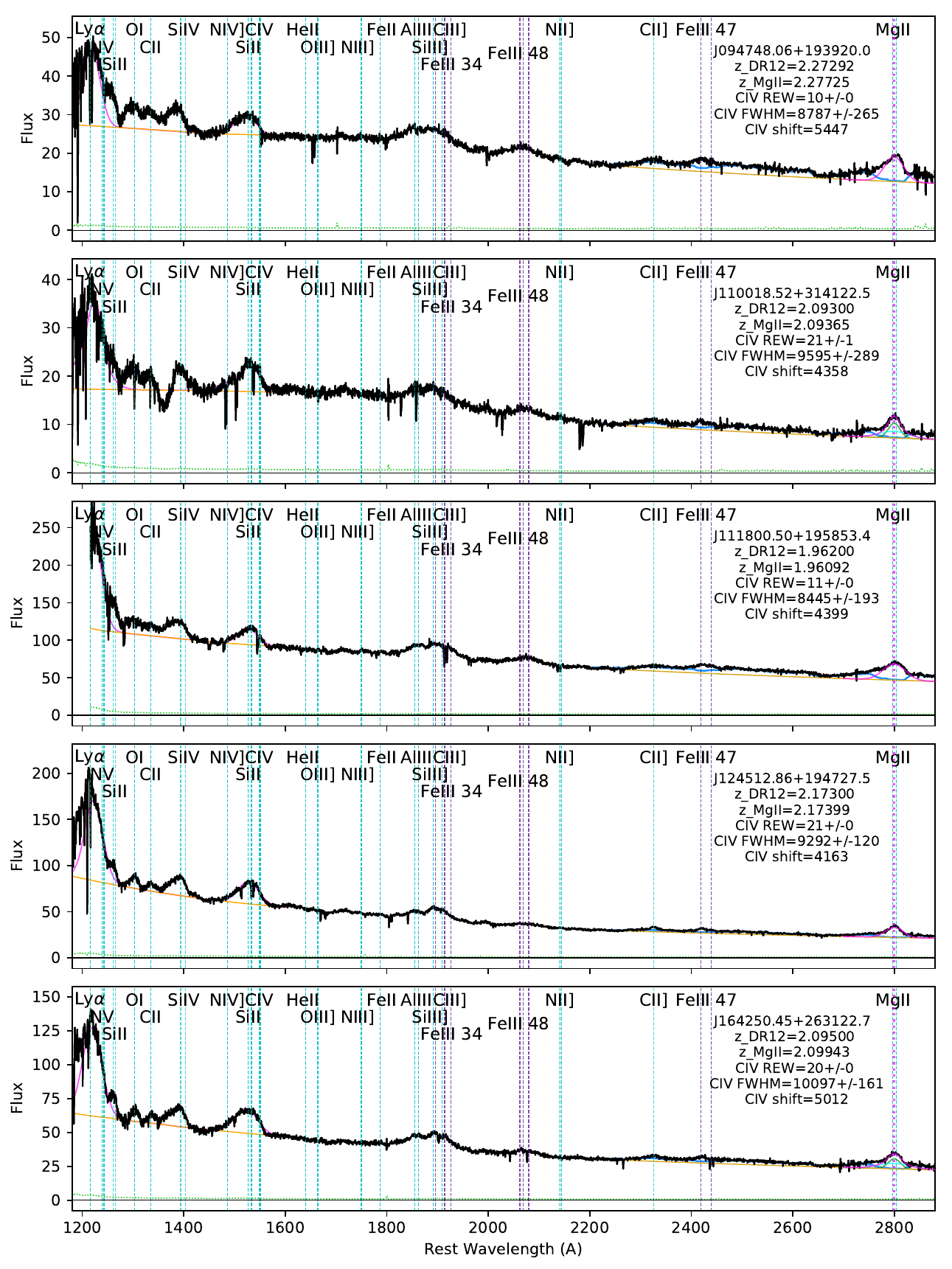}
    \caption{Selected spectra with high signal to noise, and blueshifted >4000 \kms, using the same color scheme as the spectra above.}
    \label{fig:fig_figure15}
\end{figure*}

Here we present spectra of individual quasars to demonstrate the reality of the highest blueshift cases and illustrate some of the range in the emission line properties not evident from the median spectra. Figures \ref{fig:fig_figure14} - \ref{fig:fig_figure16} are sets of individual spectra of high \civ\ blueshift quasars for illustration, or spectra with extreme properties for further discussion, chosen from regions in the top and bottom panels of Figure \ref{fig:fig_figure5}. 

Figure \ref{fig:fig_figure14} presents individual spectra with a wide range of \civ\ emission strengths, and blueshifts >2000 \kms. Quasars with strong \civ\ REW and high blueshift are rare in our sample (see Figure \ref{fig:fig_figure5}). Most quasars with extremely weak \civ\ lines have a more uniform distribution of blueshifts, from $\sim$2000 - 6000 \kms.

Figure \ref{fig:fig_figure15} presents spectra with high signal to noise and extreme blueshifts >4000 \kms. J094748.06+193920.0, has narrow absorption redshifted near \civ, and indicates \mgii\ is a possible lower limit. J110018.52+314122.5 shows a \civ\ mini-BAL, at approximately 1355\AA\ (corresponding to a gas velocity of $\sim$0.13$c$). J111800.50+195853.4 has a \mgii\ profile exhibiting an unusual blue wing, similar to the other high-ionization lines. J111800.50+195853.4 is also an example with blue asymmetry in the \mgii\ profile. J164250.45+263122.7 is a good SNR spectra with secure \civ\ blueshifts, and is somewhat ``typical'' from these highest blueshifts.  

Figure \ref{fig:fig_figure16} presents all 13 individual spectra with blueshifts >6000 \kms. J012505.27+063829.5 is the highest blueshifted case, and in spite of weak \civ\ it has a good \mgii\ profile measurement, with possible blueshifted \siv\ as additional confirmation. J084842.64+540808.2 and J140701.59+190417.9 are 2nd largest blueshifts, confirmed by multiple weak low-ion lines, and have very broad and flat \civ\ profiles. J01410.94+043210.7 has strong line measurements despite the lack of \siii/\siiv\ coverage, and coincidentally is one of the top ten highest in \civ\ blueshift. It also has an absorption spike red-ward of \civ, which could indicate \civ\ is more blueshifted, and \mgii\ may have a blue wing. 

Analysis of these individual quasars confirms the reality of trends in the extremes of Figure \ref{fig:fig_figure5}, and highlight unique features that could otherwise be lost by median composite spectra.

\section{Discussion}
\label{sec:sec_discussion}

Blueshifts observed in the broad emission lines of quasar \civ\ serve as valuable indicators of outflows in moderately to highly ionized gas from the regions emitting these broad lines (see Section \ref{sec:sec_intro}). We present blueshift measurements of \civ\ in a total of 39,249 quasars with redshifts ranging from 1.52 to 2.42, as observed in the SDSS. Our main objective is to gain insights into the nature and origins of these outflows. This is achieved by studying the characteristics of the \civ\ blueshifts and establishing connections between the blueshifts/outflows and other spectral properties of the quasars. Constraining outflow speeds necessitates reliable systemic redshifts. Notably, our approach differs from prior studies as we determine the quasar's systemic redshifts exclusively through fitting the low-ionization \mgii\ emission line, which is considered close to systemic redshift of the quasar \citep[e.g.][]{Shen+08,Shen+16b,Li+17}. This method allows us to explore a unprecedented range of outflow velocities compared to previous work, reaching blueshifts > 6000 \kms, and maintain a sample size that is statistically significant. In this section we will discuss our results for the full sample, and interesting subsamples. 

Broad emission lines offer an important advantage over absorption lines when studying outflows, because they aren't confined to gas along the direct line of sight to the quasar's emission sources. However, this flexibility also means that the gas can be moving at various angles relative to our line of sight, which complicates the relationship between observed blueshifts and outflow velocities, as it becomes contingent on the specific modeling used. The blueshifts we measure by analyzing the emission-line centroids only give us a lower threshold for the actual flow velocities. In situations where there are exclusively radial motions, and negligible additional sources of line broadening, the most accurate estimate of outflow velocities comes from the most pronounced blueshifts in the blue portion of the observed line profiles \citep[e.g., the $v_{98}$ parameter as employed by ][]{Perrotta+19}. By utilizing the parameters provided in our study, we can approximate the actual \civ\ outflow velocities based on the blueshifts of the line centers, plus half of the measured full width at half maximum (FWHM), while also accounting for half of the doublet separation in \civ\ (equivalent to 250 \kms). For example, in the high-blueshift quasar J100351+001502 shown in Figure \ref{fig:fig_figure16}, we could infer that the \civ\ outflow speed is roughly 6009 + 5296 - 250 $\sim$ 11,000 \kms. 

\begin{figure*}
	\includegraphics[width=\textwidth]{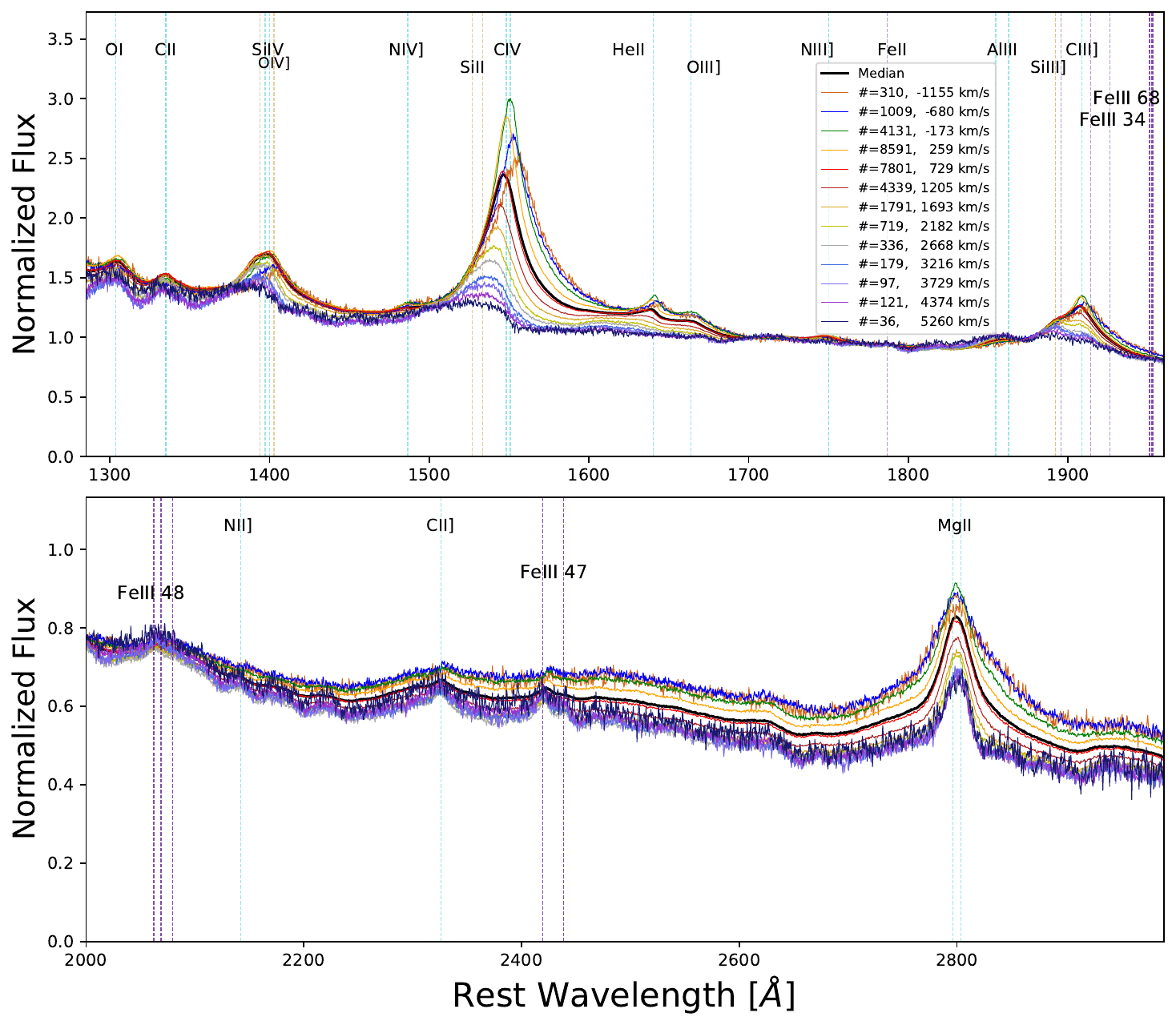}
    \caption{Median composite spectra from the Composite Sample. These spectra are generated from vertical columns, from Figure \ref{fig:fig_figure8}, defined 500\kms\ in width from $-$1500 to 6000 \kms. Color coded in the legend are the number of spectra used to generate each composite, and their median \civ\ blueshift. For more information, see Table \ref{tab:tab_table2}. Spectra are normalized to unity at $\sim 7000$\AA, before taking the median. Spectra included with \civ\ velocity over 4000 \kms\ are visually confirmed to have similar blueshifted profiles as \siiv\ and/or confirmed \mgii\ redshift with \siii. Something straight forward to correlate is stronger low-ion lines and weaker \heii\ and \ciii]/\aliii\ at larger blueshifts. }
    \label{fig:fig_figure9}
\end{figure*}

\begin{figure}
    \includegraphics[width=\columnwidth]{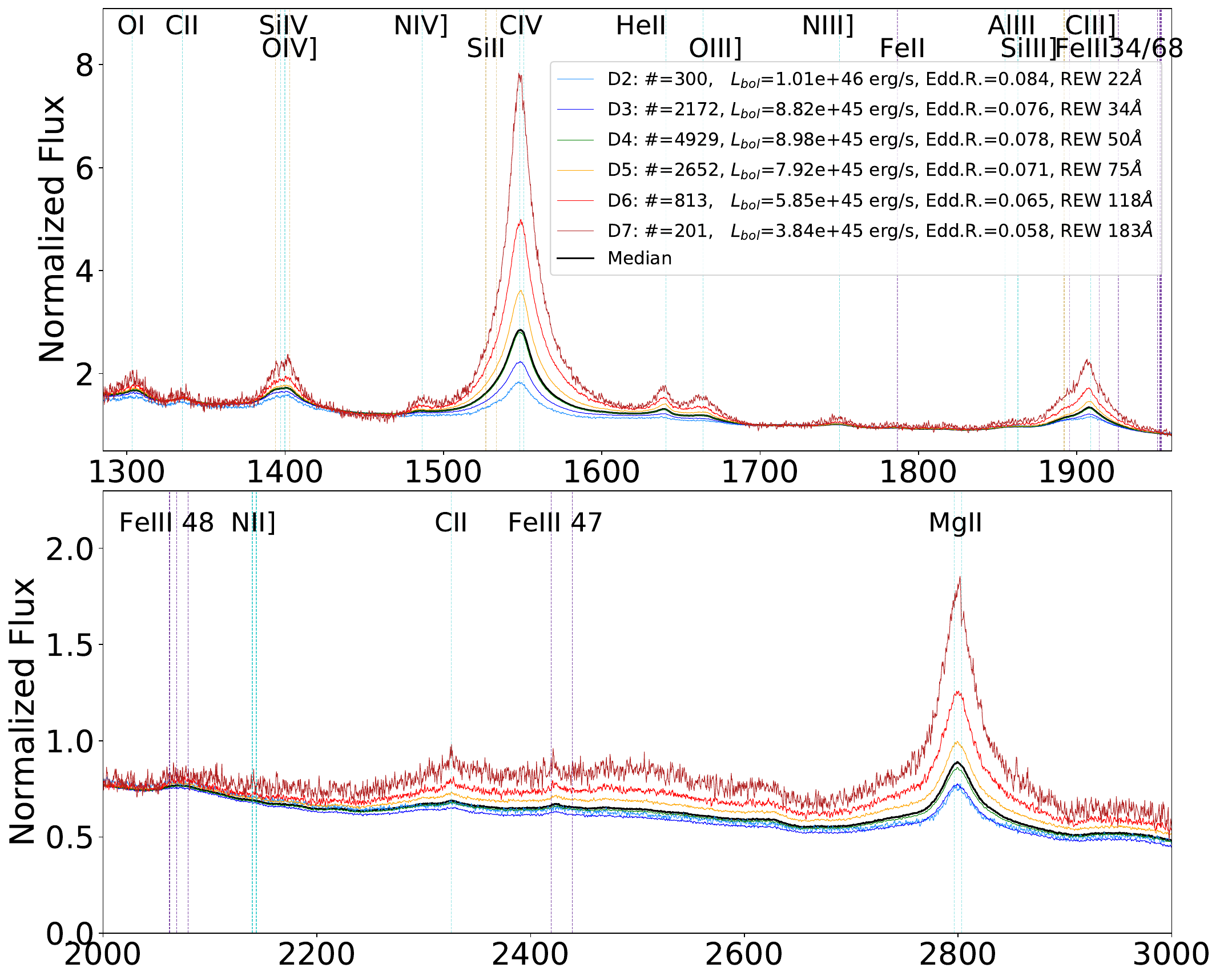}
    \includegraphics[width=\columnwidth]{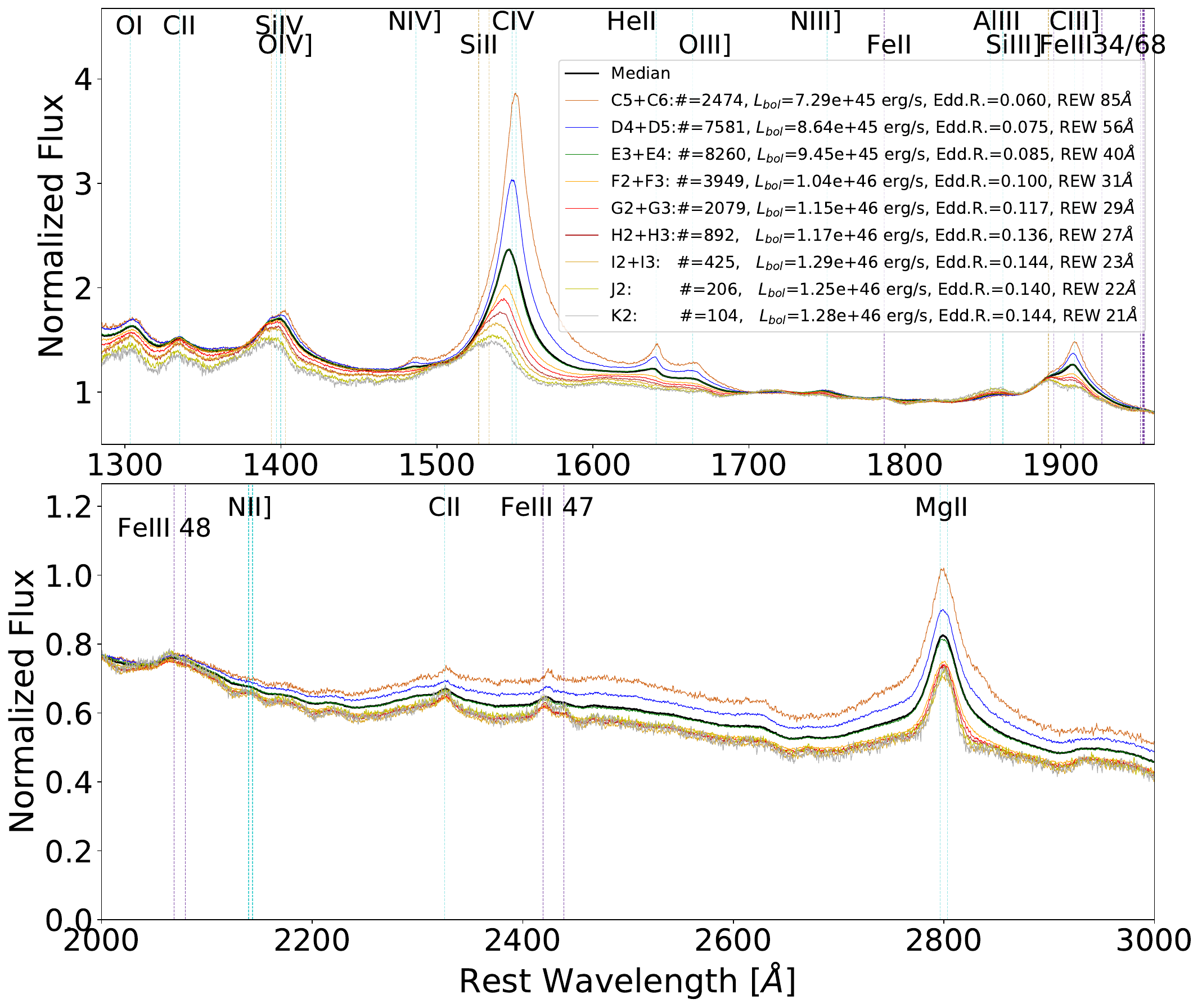}
    \caption{Median composite spectra from selected regions of Figure \ref{fig:fig_figure6}. The legend presents the number of spectra used for the composite, and median values for luminosity, Eddington Ratio, and \civ\ REW. \textit{Top:} Ranges of \civ\ REW between velocities 0 and 500 \kms. Each bin is defined moving from 16 to 251\AA, using $\sim 0.2 $ dex increments. \textit{Bottom:} Diagonal region across the population as REW decreases and \civ\ blueshift increases. }
    \label{fig:fig_figure10}
\end{figure}

Our composite median spectra reveal consistent trends of weakening \heii\ with increasing \civ\ blueshift (see Figures \ref{fig:fig_figure9} and \ref{fig:fig_figure10}). We see that high blueshifts correspond to both weaker \civ\ strength, broader emission profile, and higher Eddington ratios (see top panel of Figure \ref{fig:fig_figure12}). We are the first to show that these trends extend to the most extreme blueshift velocities, which indicate radiative line-driving may be one of the mechanisms generating extreme outflows \citep{Leighly+04a,Richards+11,Rankine+20,Temple+23}. 

Observational evidence strongly supports both of these factors as correlating with faster outflows. For instance, the inverse correlation between the \heii\ \lam 1640 emission-line REW and larger \civ\ emission-line blueshifts \citep{Leighly+04a,Richards+11,Rankine+20,Temple+23} as well as faster \civ\ broad absorption-line (BAL) outflows \citep{Baskin+13,Hamann+19,Rankine+20} indicates the importance of softer UV spectra in driving outflows. This inverse relation is also observed in hyper-luminous quasars \citep{Vietri+18}. Furthermore, larger \civ\ blueshifts have been found to correlate with higher Eddington ratios \citep{Baskin+05,Coatman+16,Rankine+20,Temple+23}. Higher Eddington ratios can provide a stronger radiative driving force relative to gravity. Weak \heii\ indicates a weaker far-UV flux, thus a softer UV continuum, and helps maintain moderate ionization levels and substantial opacities in the outflow to enable radiative driving in the near-UV \citep{Leighly04b}. 

As \civ\ REW decreases with blueshift, we see low-ionization lines becoming stronger, and a trend of weaker \ciii] relative to \aliii\ and \siiii]. There have been possible models proposed to explain this trend, but higher densities and/or lower degrees of ionization would decrease \ciii] relative to \siiii] and \aliii\ \citep{Leighly+04a,Leighly04b,Richards+11}.

\begin{table*}
\centering
\begin{tabular}{ |c|c|c|c|c|c|c|c|c|c|c| } 
\hline
Velocity & Total & \lam $L_{\lambda}$ & Eddington & REW & \civ\ & Blueshift & FWHM & REW & FWHM & $\sigma$ \\
Range & Spectra & [ergs s$^{-1}$] & Ratio & (\civ) & Blueshift & Std. Dev. & (\civ) & (\mgii) & (\mgii) & (\mgii) \\
& & & & [\AA] & [km s$^{-1}$] & [km s$^{-1}$] & [km s$^{-1}$] & [\AA] & [km s$^{-1}$] & [km s$^{-1}$] \\
\hline
$-$1500 - $-$1000 & 310 & 8.20$ \times 10^{45} $ & 0.04 & 60 & $-$1155 & 132 & 6479 & 38 & 6414 & 2838 \\
$-$1000 - $-$500 & 1009 & 7.87$ \times 10^{45} $ & 0.04 & 62 & $-$680 & 136 & 5767 & 38 & 6112 & 2735 \\
$-$500 - 0 & 4131 & 8.31$ \times 10^{45} $ & 0.06 & 61 & $-$173 & 139 & 4526 & 37 & 5254 & 2448 \\
0 - 500 & 8591 & 8.65$ \times 10^{45} $ & 0.08 & 52 & 259 & 143 & 4335 & 35 & 4776 & 2243 \\
500 - 1000 & 7801 & 9.05$ \times 10^{45} $ & 0.08 & 41 & 729 & 142 & 4927 & 33 & 4659 & 2155 \\
1000 - 1500 & 4339 & 9.48$ \times 10^{45} $ & 0.09 & 34 & 1205 & 142 & 5473 & 31 & 4425 & 2072 \\
1500 - 2000 & 1791 & 1.00$ \times 10^{46} $ & 0.10 & 30 & 1693 & 138 & 5967 & 28 & 4161 & 1970 \\
2000 - 2500 & 719 & 9.63$ \times 10^{45} $ & 0.12 & 27 & 2182 & 143 & 6257 & 26 & 3822 & 1843 \\
2500 - 3000 & 336 & 1.04$ \times 10^{46} $ & 0.13 & 23 & 2668 & 142 & 6685 & 25 & 3781 & 1789 \\
3000 - 3500 & 179 & 1.00$ \times 10^{46} $ & 0.13 & 20 & 3216 & 146 & 7447 & 24 & 3680 & 1769 \\
3500 - 4000 & 97 & 9.34$ \times 10^{45} $ & 0.14 & 18 & 3729 & 148 & 8087 & 23 & 3686 & 1725 \\
4000 - 5000 & 121 & 9.86$ \times 10^{45} $ & 0.12 & 14 & 4374 & 274 & 8951 & 24 & 3908 & 1860 \\
5000 - 6000 & 36 & 8.86$ \times 10^{45} $ & 0.11 & 10 & 5260 & 295 & 9436 & 21 & 4018 & 1942 \\
$-$1500 - 6000 & 29460 & 8.94$ \times 10^{45} $ & 0.08 & 43 & 538 & 804 & 5048 & 33 & 4682 & 2195 \\
\hline
\end{tabular}
\caption{Median line properties at different velocity bins of the Composite Sample in Table \ref{tab:tab_table1}, with a limited luminosity range. These bins correspond to the spectra in Figure \ref{fig:fig_figure9} \& the top panel of Figure \ref{fig:fig_figure10}. Medians were taken of the distribution of values in each bin.}
\label{tab:tab_table2}
\end{table*}

\subsection{Eddington Ratio \& Color Distributions}
\label{sec:sec_other_distributions}

\begin{figure}
    \includegraphics[width=\columnwidth]{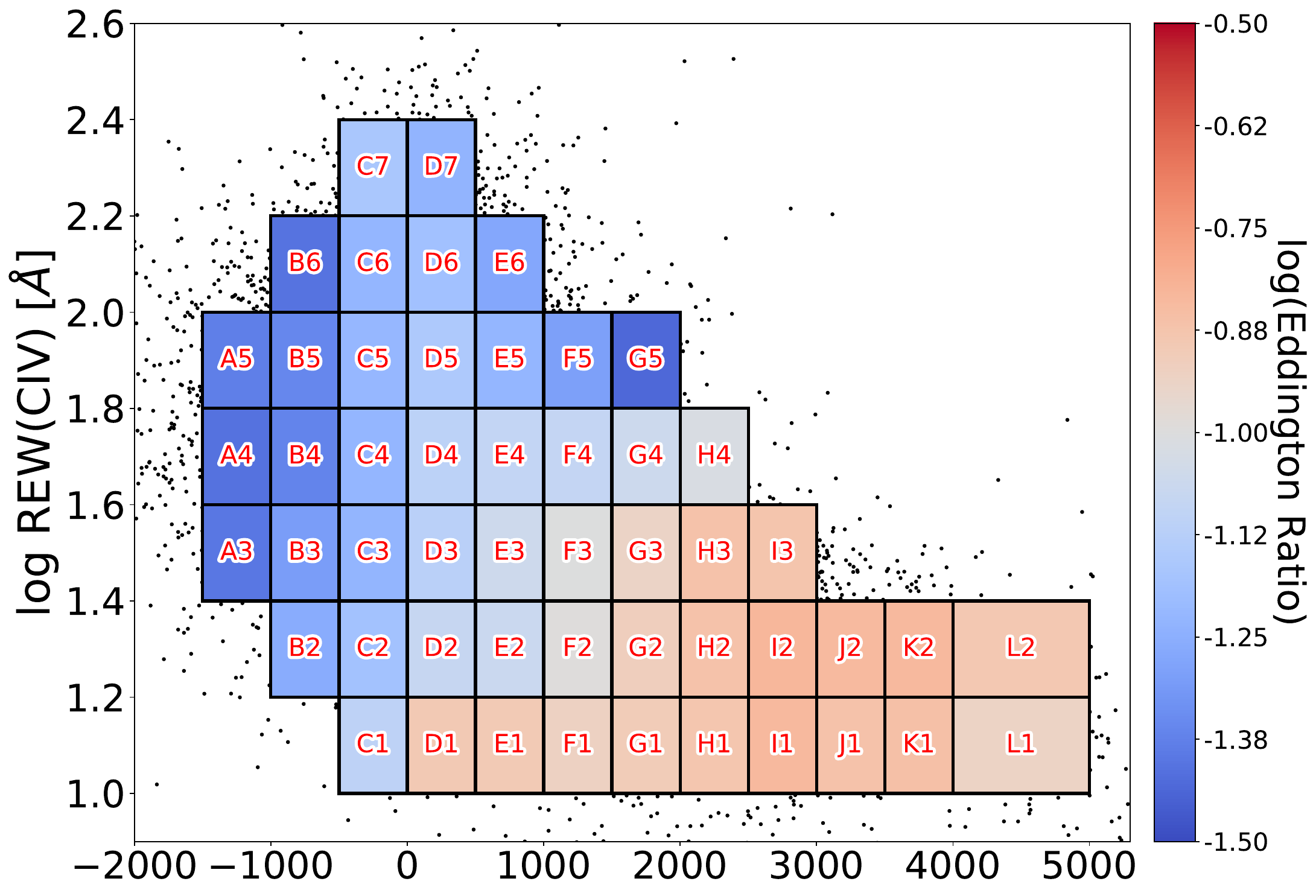}
    \includegraphics[width=\columnwidth]{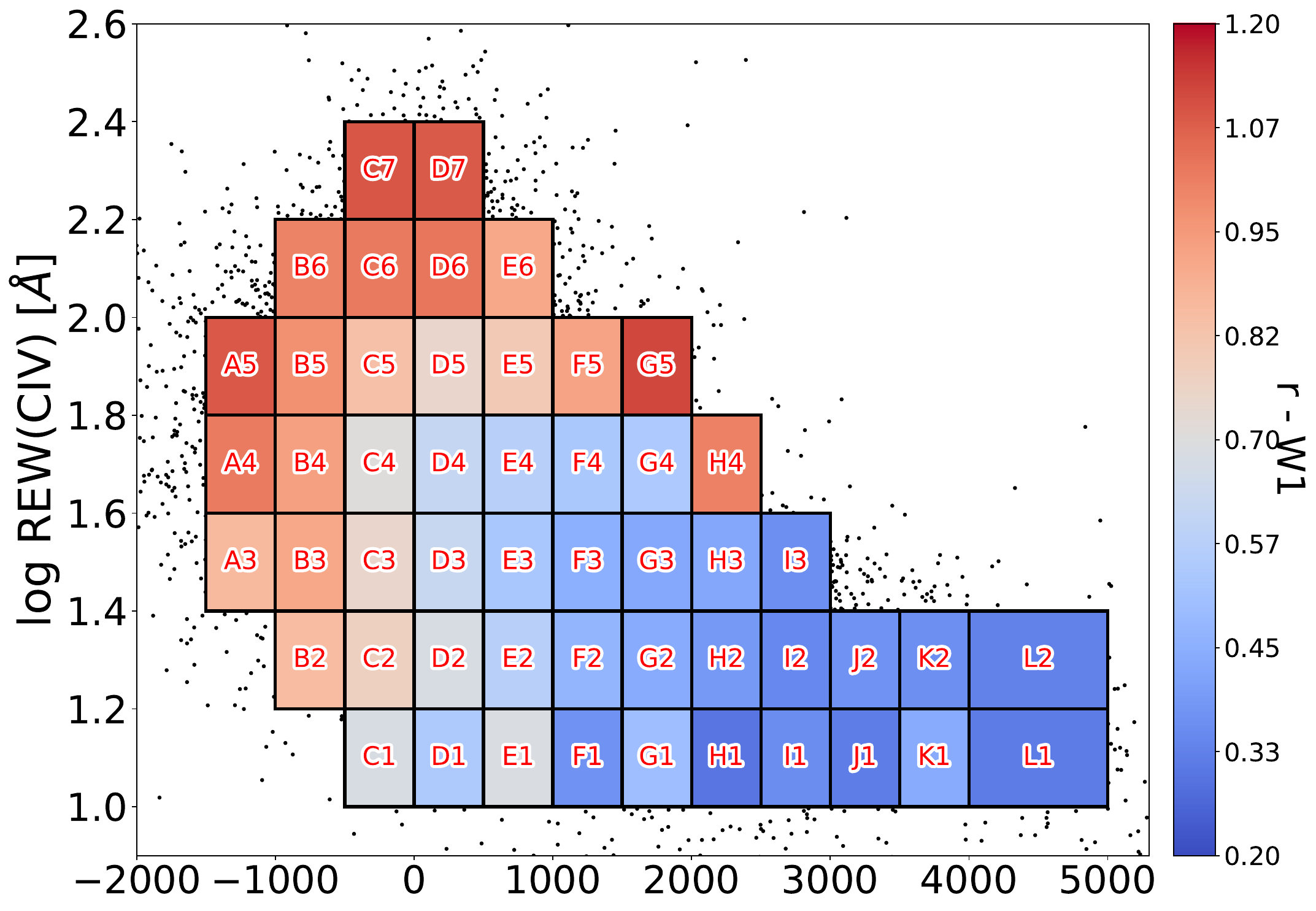}
	\includegraphics[width=\columnwidth]{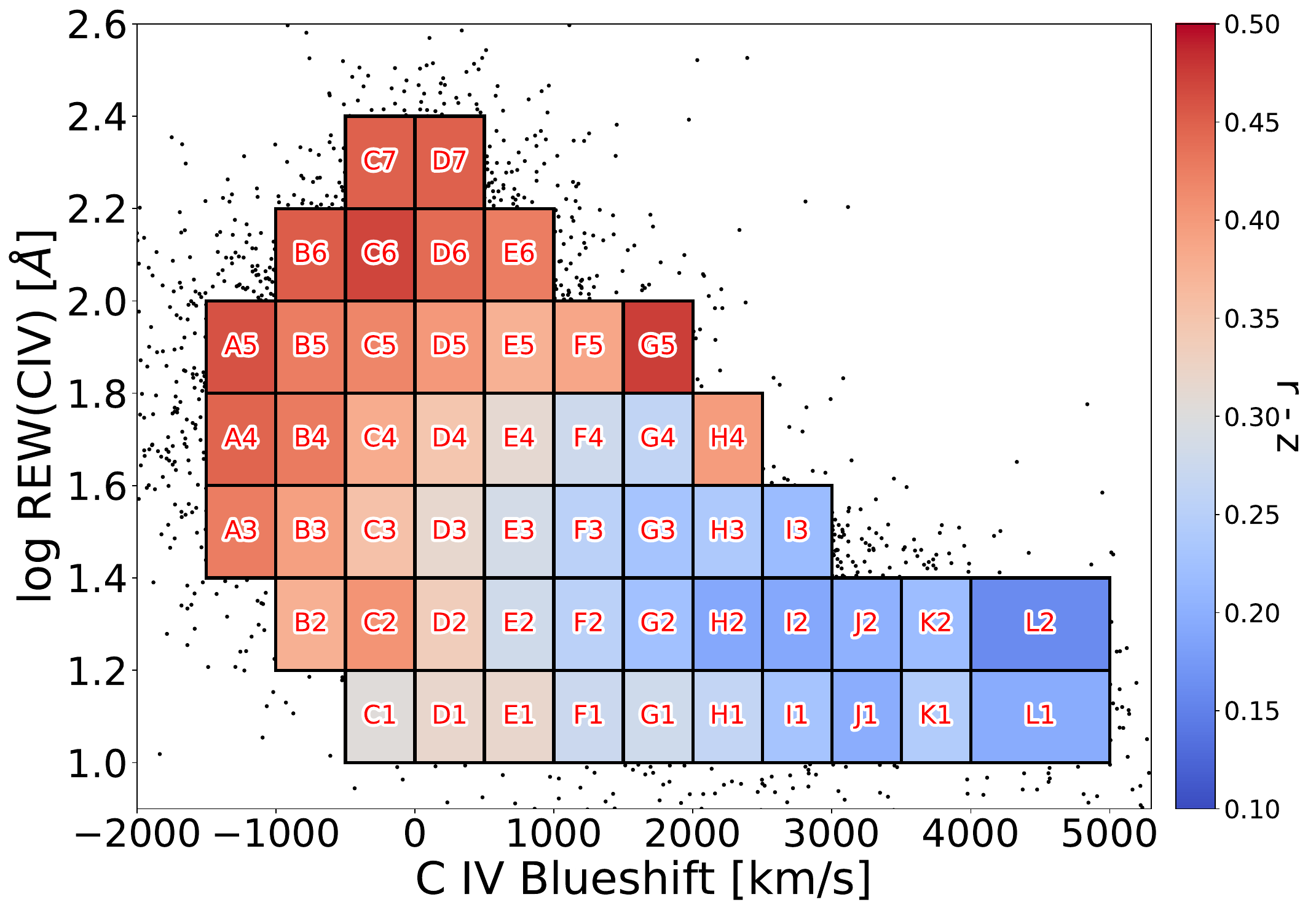}
    \caption{REW(\civ) vs \civ\ velocity shift with the same regions as in Figure \ref{fig:fig_figure7}, color coded for various parameters within the respective region. \textit{Top:} color coded for median Eddington ratio from the BH Mass sample of Table \ref{tab:tab_table1}. \textit{Middle:} color coded with median \rmwo from spectra that have $W1$ detection (see Color Sample in Table \ref{tab:tab_table1}).
    \textit{Bottom:} color coded for \rmz. 
    }
    \label{fig:fig_figure12}
\end{figure}

Figure \ref{fig:fig_figure12} displays several trends which reveal complexities to our goal of finding what drives powerful outflows. We find larger Eddington ratios also correlates with increasing blueshift \citep[bottom panel of Figure \ref{fig:fig_figure12}, and also indicated in][]{Baskin+05,Coatman+16,Rankine+20,Temple+23}. When the luminosity of the quasar is near to \ledd, and the gravitational force from the SMBH is nearly balanced, it becomes easier to drive an outflow. Eddington  ratio may therefore be an important factor in allowing an ensemble of conditions (e.g., high luminosity, continuum shape) to take maximum affect. 

\civ\ blueshift is more strongely correlated to Eddington ratio than $L_\text{bol}$. This correlation is evident in comparing Figure \ref{fig:fig_figure12} to \ref{fig:fig_figure7}, and from our composites contributing to Figure \ref{fig:fig_figure9}, where we find a strong trend with Eddington ratio after sorting by blueshift at roughly fixed $L_\text{bol}$ (see Table \ref{tab:tab_table2}).

Larger Eddington Ratio and softer far-UV spectra are probably the fundamental physical drivers behind faster, more powerful BLR outflows, and both appear to correlate with larger blueshifts in out data. While different combinations of these factors might produce the observed blueshifts in different individual quasars, the overall trends evident from our composite samples (Figure \ref{fig:fig_figure9} and Table \ref{tab:tab_table2}) suggests that both larger Eddington ratio and softer far-UV spectra are involved in producing the largest blueshifts.

We confirmed any reddening that could change the Eddington ratios in Figure \ref{fig:fig_figure12} may be influenced by about A$_{3000}$ = 0.57 magnitudes extinction, or a factor of 1.69 flux suppression. The effect on Eddington ratios will go as the square-root of that, a factor of 1.30 or 0.11 in log. Comparing that to the bottom panel in Figure \ref{fig:fig_figure12} shows that performing a reddening correction as above would weaken the trend but not undo it. This is a maximal correction, because the 0.3 color difference is generous. More importantly, we assumed all of the color difference is due to dust reddening instead of intrinsic color differences in the quasars that might be related to outflows/blueshifts. A reddening correction, if warranted, does not undo these trends. Additional correlations with SED shape can also influence outflow velocities \citep{Richards+11,Baskin+13}.

\begin{figure}
	\includegraphics[width=\columnwidth]{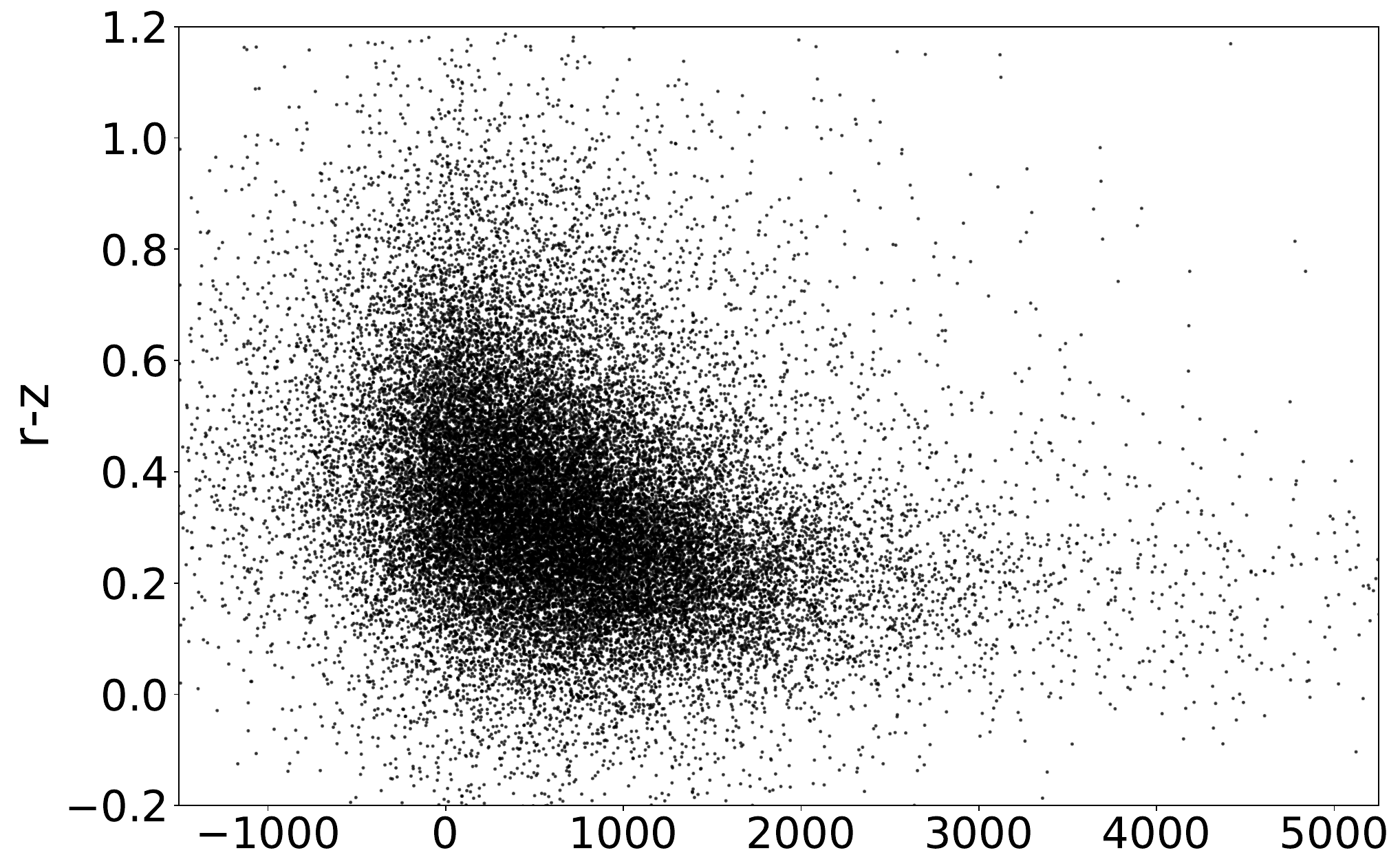}
    \includegraphics[width=\columnwidth]{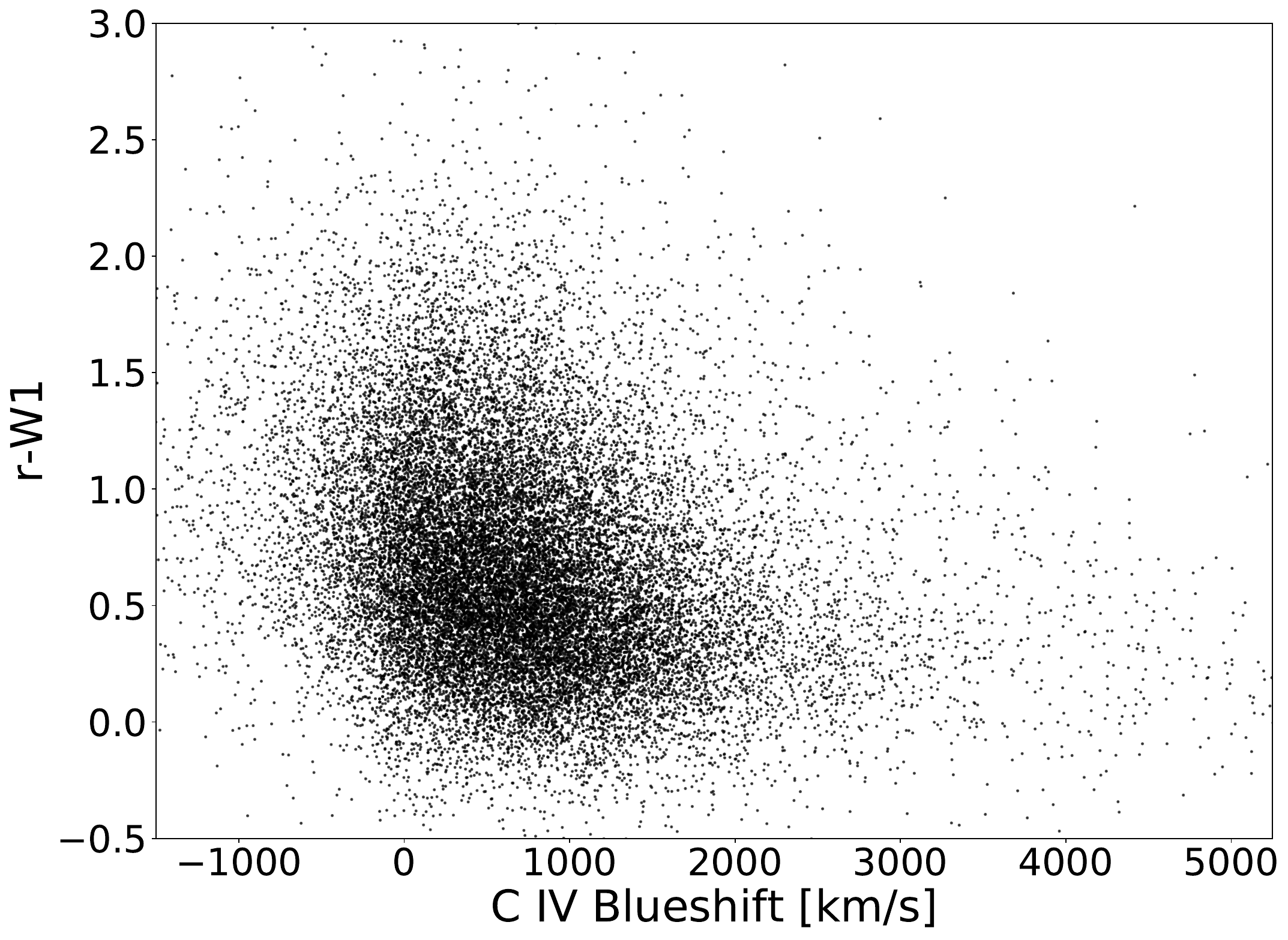}
    \caption{\civ\ velocity shift vs color cuts for quasars with $r$,$z$, and/or $W1$ detections. \textit{Top:} Blueshift vs (\rmz) color in our sample with $r$ and $z$ mag data. \textit{Bottom:} Blueshift vs (\rmwo) color in our $W1$-detected quasars.}
    \label{fig:fig_figure13}
\end{figure}

We see trends in UV color (see Figures \ref{fig:fig_figure13} and \ref{fig:fig_figure12}), showing that bluer color across the UV to Near-IR correlates with blueshift out to extreme velocities. It has been shown that extreme reddening across UV to IR can also correlate with extreme blueshifts/outflows \cite{Perrotta+19}. The ``less luminous'' quasars are redder. In fact, their luminosities are underestimated because of reddening, and the Baldwin Effect is at least partly a reddening trend, and luminosity. Color as measured by spectral index, or broadband color, have indicated slight asymmetric blueshifts in \civ\ correlated with bluer colors \citep{Richards03,Leighly+04a,Gallagher+05,Richards+11}.  

Quasars can have intrinsically red UV continuum without involving dust. Quasars defined by a steep optical-continuum, which are intrinsically red without dust reddening, have been seen to have stronger \civ, \ciii], and \heii\ emission \citep{Richards03}. Redder colors can also be attributed to more extinction/reddening along inclined viewing angles in \citep{Baskin+13}, but they do not see a color trend for higher BAL speeds, which should correlate with inclination angle. Emission line blueshifts should have weaker dependence on viewing angle because they sample material moving in all directions, not serendipitous absorption along line of sight. Small differences in reddening could be small differences in viewing angle, therefore extremely small changes in cos($\theta$) factor affecting the measured blueshifts. A more detailed analysis would require specific models for the dust and outflow geometries around quasars, but is beyond the scope of this work. We note that the trends seen in Figures \ref{fig:fig_figure13} and \ref{fig:fig_figure12} could be due mostly, or entirely, to intrinsic color differences in the quasar SEDs.

\begin{figure}
    \includegraphics[width=\columnwidth]{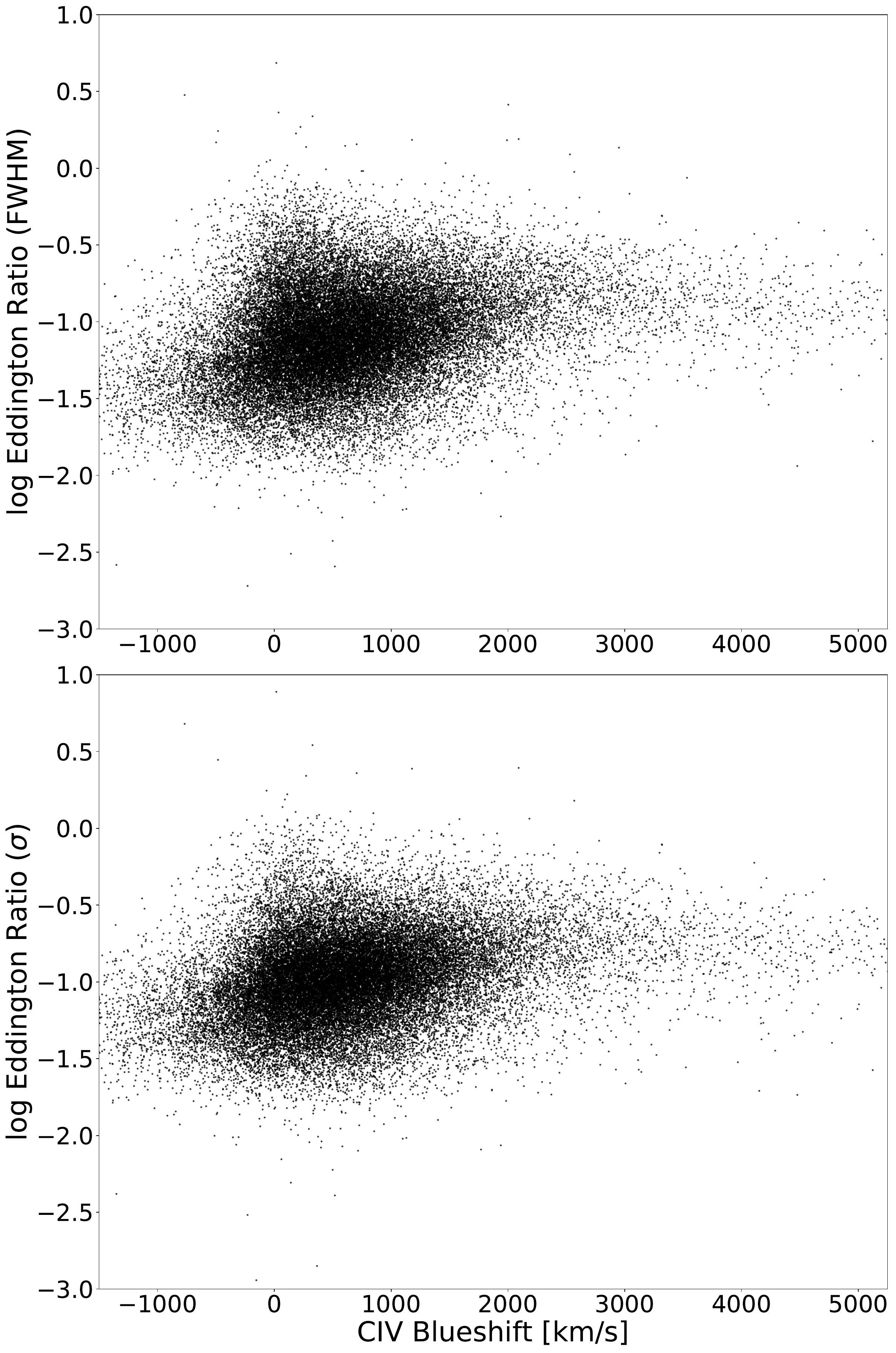}
    \caption{Distribution of \civ\ Blueshift vs Eddington ratio using $\sigma$ or FWHM of \mgii\ for computing $L_{\text{Edd}}$. \textit{Top:} Eddington ratio computed with FWHM of \mgii. \textit{Bottom:} Eddington ratio computed with $\sigma$ \mgii. Eddington luminosity computed using $\sigma$ shows a narrower distribution than with FWHM. }
    \label{fig:fig_figure11}
\end{figure}

Trends with Eddington ratio and color both appear along the curved diagonals (upper left to lower right) in Figure \ref{fig:fig_figure12} illustrate how \civ\ blueshift, and  thus outflow speed, is a multi-dimensional problem. This complexity is not evident from simpler 2-parameter plots like Figures \ref{fig:fig_figure11} and \ref{fig:fig_figure13}. Blueshifts depend simultaneously on several quasar observables, including the \civ\ REW and near-UV continuum color, in addition to the more fundamental physical drivers like Eddington ratio and far-UV spectral shape (indicated by \heii). These parameters might be connected to each other in different ways. For example, photoionization models of quasar BLRs have shown that softer ionizing spectra naturally produce smaller emission-line REWs \citep{Korista+97,Korista+98}, and might explain why these parameters are both related to larger blueshifts. However, more work is needed to understand if softer ionizing spectra are sufficient, or even necessary, to explain the strong observed trend between \civ\ REW and blueshift. Another possibility is that the observed near-UV colors are tied to reddening by dust along our line of sight and perhaps our viewing perspective of the accretion disc and outflow environment \citep{Richards+02}.

Overall, an interesting result that is revealed by our color coded ``3-dimentional'' parameter space of \civ\ REW vs blueshift is that they show trends better than the 2D scatter plots. This reveals that the blueshifts are a multi-dimensional/multi-parameter problem. Future studies may look for where large blueshifts cluster in a multi-dimensional parameter space to determine their fundamental causes.

\subsection{Individual Spectra}
\label{sec:sec_discuss_individual_spectra}

Broad studies of large quasar populations seldom present rare and interesting cases such as spectra with the most extreme properties. We present individual spectra to convey two points. i) demonstrate the reality of the most extreme cases with blueshifts >6000 \kms. Our composite figures only consider blueshifts <5000 \kms\ because there are too few to construct meaningful medians, and the wide range in blueshifts would blend the \civ\ profile. ii) display well-measured and interesting spectra at large blueshift. For example, spectra with extreme properties of \civ\ like large REW, and cases of a strong blue wing in \mgii, that may not be evident from composite spectra.

A surprising feature in many high-blueshift spectra is a peculiar narrow absorption spike at rest wavelengths of \civ. This absorption feature is not typical in quasars with average broad and symmetric \civ\ profiles, and is smoothed out of median composite spectra. Five of our top 13 fastest-blueshift spectra have this narrow absorption feature (e.g., J012505.7+063829.5, J014010.94+043210.7, and J105208.27+151004.0). This absorption feature is commonly seen in spectra with blueshifts >4000 \kms, and also high signal to noise spectra when \civ\ is broad and blueshifted (see Figure \ref{fig:fig_figure16}). An origin of this absorption spike is not immediately clear. One hypothesized explanation is a viewing perspective effect of the accretion disk and BLR, which forms a narrow associated absorption-line system at systemic redshift \citep{Richards+21}. 

Another unexpected result is blueshifted wings in \mgii\ profiles. Outflows occur only, or primarily, in the high-ionization lines of the BLR. Blueshifted wings are typically observed in these high-ionization line profiles, not low-ions, and is a key feature in using low-ionization lines for indicating systemic redshift. An occasional, weak blue wing in \mgii\ is seen in spectra with the most blueshifted \civ, and indicates that a component of even the low-ion gas is involved in fast outflows. J01410.94+043210.7 has strong line measurements that show a blue excess in the \mgii\ profile, and also contains a \civ\ absorption spike at a slightly redshifted wavelength, which could also indicate our blueshift velocity is a lower limit. There is a similar scenario in J094748.06+193920.0. It has been speculated that low-ionization broad-lines could also be involved in outflows under extreme conditions \citep{Hamann+17,Gillette+23b}. These blueshifted wings may have consequences for \mgii\ mass estimates of the SMBH, or for other measurements that assume the broad-lines are virialized \citep{Shen+16a,Shen+16b,Coatman+16,Shen+19,Rankine+20,Temple+23}. Future studies may constrain velocities for these extremely blueshifted outflows, perhaps with non-UV emission lines.

\begin{figure*}
 	\centering
 	\includegraphics[width=0.95\textwidth]{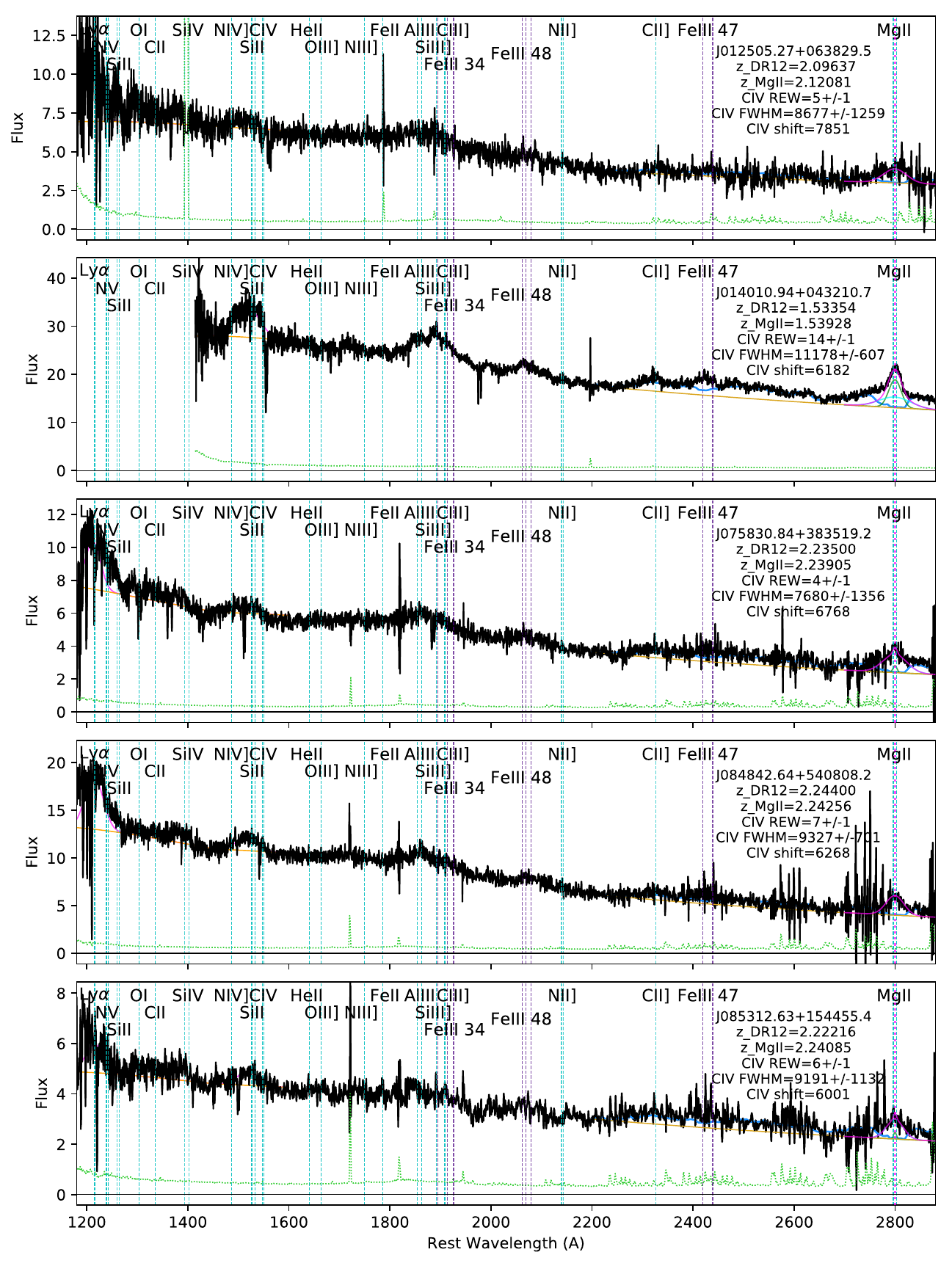}
    \caption{Selected spectra with highest \civ\ blueshifts > $\sim$6000 \kms, using the same color scheme as the spectra above.}
    \label{fig:fig_figure16}
\end{figure*}
\begin{figure*}
 	\centering
 	\includegraphics[width=0.95\textwidth]{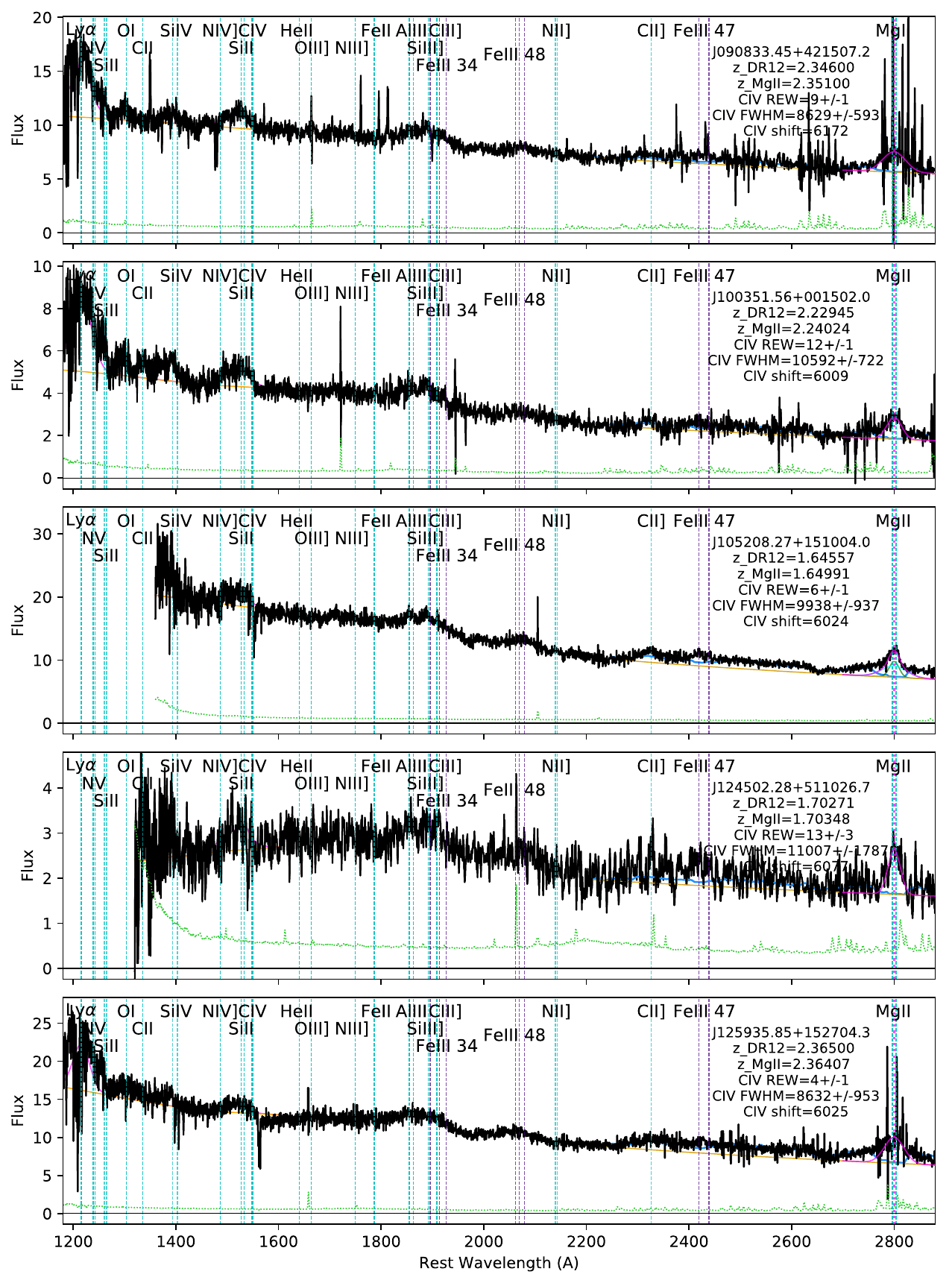}
    \raggedright \contcaption{.}
\end{figure*}
\begin{figure*}
 	\centering
 	\includegraphics[width=0.95\textwidth]{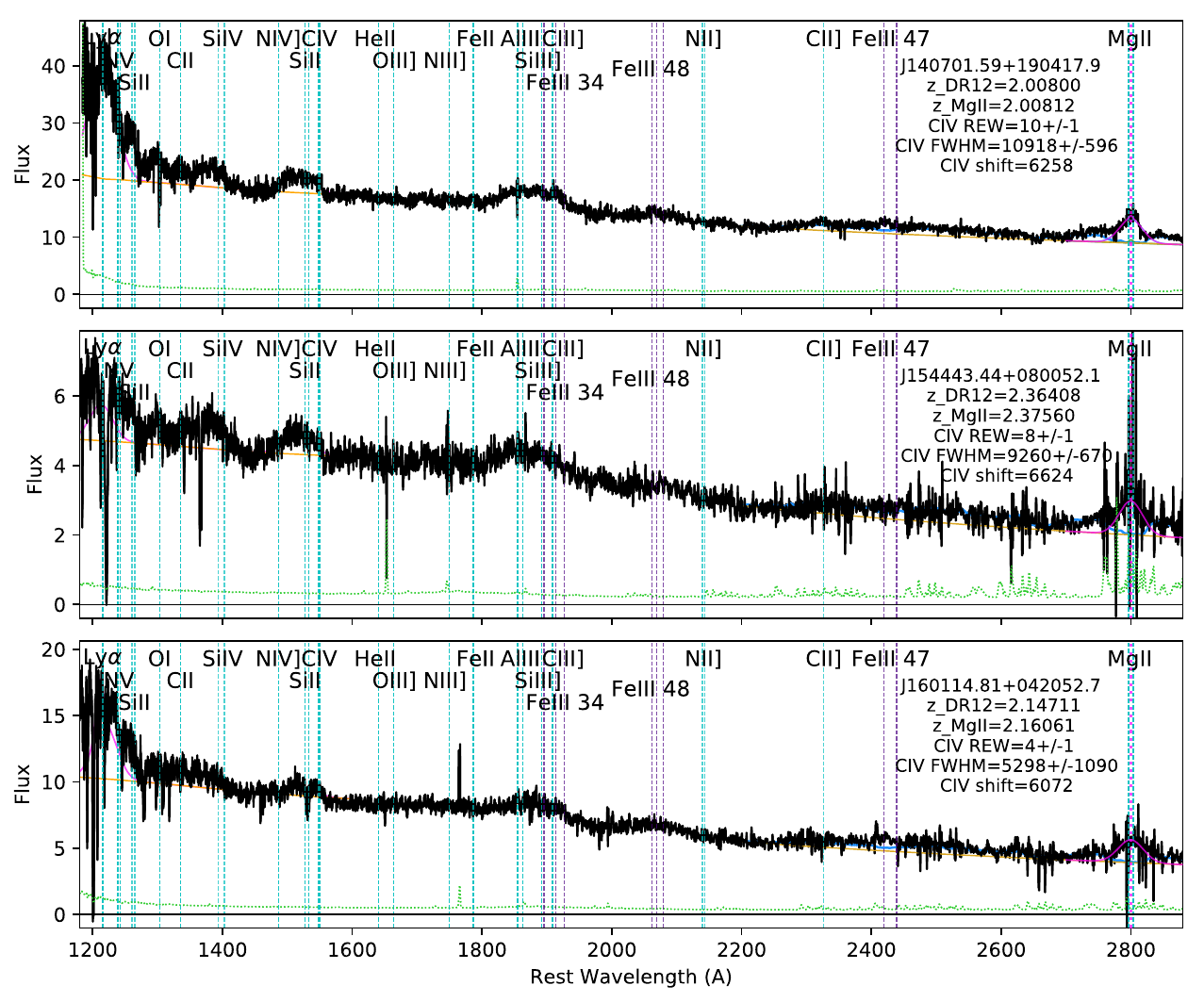}
    \raggedright \contcaption{.}
\end{figure*}

\section{Summary}
\label{sec:sec_summary}

We present a sample of 39,249 quasars with improved redshift estimates to constrain outflows at extreme speeds, to understand the factors driving outflows across a wide range of velocities. We focus on the blueshift velocities of \civ\ emission, relative to \mgii, and comparing them with other spectral properties.

This study differs from previous work by determining redshifts solely from the \mgii\ line, allowing us to explore an unprecedented range of outflow velocities and maintain a statistically significant sample size. We identify two primary factors contributing to stronger outflows: higher Eddington ratios, which provide a stronger radiative driving force relative to gravity, and a softer far-UV continuum, where weaker far-UV flux helps maintain moderate ionization levels and substantial opacities in the outflow.

Our analysis reveals consistent trends, such as the weakening of \heii\ emission with increasing \civ\ blueshift. High blueshifts correspond to weaker \civ\ strength and broader emission profiles. This study suggests that radiative line-driving may be a mechanism generating extreme outflows. As the \civ\ rest equivalent width (REW) decreases with blueshift, low-ionization lines become stronger, and there is a trend of weaker \ciii] relative to \aliii\ and \siiii]. A possible explanations for this trend is higher densities, and/or lower degrees of ionization, at farther BLR distances. 

We also find a correlation between larger Eddington ratios and increasing blueshift, and discuss the potential influence of reddening on the Eddington ratios but note that a reddening correction would not undo the observed trends.

Additionally, correlations with SED shape and color are explored. Asymmetric blueshifts in \civ\ are correlated with bluer continuum colors in the UV, particularly in less luminous quasars. Quasars can have intrinsically redder SEDs without involving dust, and the trends observed in continuum shape and reddening may be attributed to intrinsic color differences in the quasar SEDs.

Finally we look at individual spectra with extreme characteristics in their \civ\ emission profiles. We confirm the reality of trends measured across the breadth of \civ\ parameter space, including line strength, profile width, and blueshift.

Our analysis suggests that the relationship between \civ\ REW and blueshift is better understood in a multi-dimensional parameter space rather than through 2D scatter plots, indicating that blueshifts are influenced by multiple factors. We propose that future studies should investigate where large blueshifts cluster in multi-dimensional parameter space to determine the fundamental causes.

\section*{Acknowledgements}

JG and FH acknowledge support from the USA National Science Foundation grant AST-1911066. JG thanks G. Richards for insightful discussion, and Y. Shen for providing emission templates in personal correspondence. 



\bibliographystyle{mnras}
\bibliography{ms}


\bsp	
\label{lastpage}
\end{document}